\documentclass[12pt]{article}
\usepackage{amsmath}

\usepackage{graphicx}

\def\papertitlepage{\baselineskip 3.5ex\thispagestyle{empty}}
\def\preprinumber#1#2{\hfill\begin{minipage}{4.2cm} #1
        \par\noindent #2 \end{minipage}}
\allowdisplaybreaks[3]

\begin{document}

\papertitlepage
\setcounter{page}{0}
\preprinumber{KEK-TH-1356}{}
\baselineskip 0.8cm
\vspace*{2.0cm}

\begin{center}
{\Large\bf Boltzmann equation in de Sitter space}
\end{center}
\vskip 4ex
\baselineskip 1.0 cm

\begin{center}

Hiroyuki K{\sc itamoto}$^{2)}$
\footnote{E-mail address: kitamoto@post.kek.jp}
and
Yoshihisa K{\sc itazawa}$^{1),2)}$
\footnote{E-mail address: kitazawa@post.kek.jp}\\
\vspace{5mm}
$^{1)}$
{\it KEK Theory Center}\\
{\it Tsukuba, Ibaraki 305-0801, Japan}\\
$^{2)}$
{\it The Graduate University for Advanced Studies (Sokendai)}\\
{\it Department of Particle and Nuclear Physics}\\
{\it Tsukuba, Ibaraki 305-0801, Japan}\\
\end{center}

\vskip 5ex
\baselineskip = 3.5 ex

\begin{center}{\bf Abstract}\end{center}

\hspace{0.7cm}
In a time dependent background like de Sitter space, Feynman-Dyson perturbation theory breaks
down due to infra-red divergences. We investigate an interacting scalar field theory in Schwinger-Keldysh
formalism. We derive a Boltzmann equation from a Schwinger-Dyson equation
inside the cosmological horizon. Our solution shows that the particle
production is compensated by the reduction of the on-shell states due to unitarity.
Although the degrees of freedom inside the horizon
leads to a small and diminishing screening effect of the cosmological constant,
there is a growing screening effect from those outside
the horizon.

\vspace*{\fill}
\noindent
April 2010

\newpage

\section{Introduction}
\setcounter{equation}{0}

\hspace{0.7cm}
Investigating field theory in de Sitter space may illuminates deep mysteries surrounding
inflation in the early universe and dark energy of the present universe.
The past and current exponential expansions of the universe are likely to be driven by 
the effective cosmological constants
of the order of GUT and neutrino mass scales respectively. We are perplexed by the huge disparity of the 
relevant energy scales.
Of course we do not understand why they are small in comparison to the Planck scale in the first place.
Phenomenologically it appears that the cosmological constant has evolved with time.
Although we may parametrize it by a scalar field with a suitable potential, its
microscopic understanding is totally lacking. 

Slow roll inflation models possess approximate conformal invariance and the conformal
symmetry plays an important role to understand the magnitude of the correlators 
\cite{Mald}\cite{Weinberg}\cite{Komatsu} and 
possible dS/CFT correspondence
\cite{Witten}\cite{Strm}\cite{BMS} .
In string theory, there seems to be no stable de Sitter vacuum as we need to consider
brane-antibrane systems to realize it.

Since our understanding is so sparse, we wonder if we are entering a completely new territory.
It might very well be the case since the standard Feynman-Dyson perturbation theory breaks down
in a time dependent background like de Sitter space. Feynman-Dyson formalism is the backbone not only
in relativistic field theory but also in classical statistical mechanics and critical phenomena.
In this sense our expertise might be confined in equilibrium physics while
our problem belongs to non-equilibrium physics.

In fact we need to use Schwinger-Keldysh formalism to investigate field theory in a time dependent background
like de Sitter space. We can derive a Boltzmann equation in this formalism which is a
standard tool to investigate non-equilibrium physics \cite{Polyakov}.
In such a setting, it is in principle possible that the effective cosmological constant changes with time.
In other words the dynamics may explain deep mysteries of this century if we can demonstrate that the 
cosmological constant decreases with time in an interacting field theory.
Although there exist several proposals along this line of thoughts in the literature, our understanding is
still in a preliminary stage \cite{Polyakov}\cite{Polyakov1}\cite{Jackiw}\cite{TW}\cite{GB}.

There is a long history of studying Boltzmann equations in Schwinger-Keldysh formalism \cite{SW}
starting from Kadanoff-Baym \cite{KB}\cite{Kd}\cite{kita} .
In this paper we derive a Boltzmann equation in de Sitter space from a Schwinger-Dyson equation.
This problem has been studied to the leading order of the derivative expansion of the Moyal product 
in the Wigner representation \cite{Hohenegger}. 
However only the energy conserving process has been identified in such a limit.
We go beyond the leading order of the expansion
to investigate the particle production effects due to energy non-conservation in de Sitter space.
We also investigate the energy-momentum tensor of an interacting scalar field theory
to estimate the effective cosmological constant.

The organization of this paper is as follows.
In section 2, we introduce a scalar field theory in de Sitter space. In section 3, we recall Schwinger-Keldysh
formalism. In section 4, we determine the full propagator inside the cosmological horizon to the leading order
in perturbation theory.
In section 5, we estimate the effective cosmological constant from the energy-momentum tensor
of a scalar field. We conclude in section 6 with discussions.

\section{Scalar field theory in de Sitter space}
\setcounter{equation}{0}

\hspace{0.7cm}
In the Poincar\'{e} coordinate, the metric in de Sitter(dS) space is
\begin{equation}\begin{split}
ds^2&=-dt^2+e^{2Ht}d{\bf x}^2\\
&=\frac{-d\tau^2+d{\bf x}^2}{H^2\tau^2} , 
\end{split}\end{equation}
where the dimension of dS space is taken as $D=4$. $H$ is the Hubble constant
and the conformal time $\tau$ is related to the cosmic time $t$ as $\tau\equiv-\frac{1}{H}e^{-Ht}$.
It assumes the value in the range $-\infty < \tau <0$ and it increases with
cosmic evolution.

In this paper, we consider a massless scalar field $\varphi$ which is minimally coupled to the dS background.
The quadratic action for the matter field is
\begin{equation}\begin{split}
S_{matter}=&\frac{1}{2}\int\sqrt{-g}d^4x\ [-g^{\mu\nu}\partial_\mu\varphi\partial_\nu\varphi] .
\end{split}\end{equation}
We find it convenient to redefine the scalar field as follows $\varphi\rightarrow H\tau\varphi$.
We can simply scale it back to find the original scalar field.  
In terms of the rescaled field, the quadratic action becomes
\begin{equation}
S_{matter}=\frac{1}{2}\int d^4x\ \varphi\left(-\partial_\tau^2+\partial_{{\bf x}}^2+\frac{2}{\tau^2}\right)\varphi .
\end{equation}
The positive frequency solution of the equation of motion with respect to the above action is
\begin{equation}
\phi_{{\bf p}}(x)=\frac{1}{\sqrt{2p}}(1-i\frac{1}{p\tau})\ e^{-ip\tau+i{\bf p}\cdot{\bf x}} ,
\end{equation}
where $p=|{\bf p}|$.
We expand the scalar field as
\begin{equation}
\varphi (x) = \int \frac{d^3p}{(2\pi)^3}\left( a_{{\bf p}}\phi_{{\bf p}}(x)
+ a_{{\bf p}}^{\dagger}\phi_{{\bf p}}^*(x)\right) .
\end{equation}
We consider the Bunch-Davies vacuum $|0\rangle$ which is annihilated by all the destruction operators 
$\forall a_{{\bf p}}|0\rangle=0$.
The propagator in such a vacuum is 
\begin{equation}\begin{split}
\langle\varphi(x_1)\varphi(x_2)\rangle
=&\int \frac{d^3p}{(2\pi)^3}\ \phi_{{\bf p}}(x_1)\phi_{{\bf p}}^*(x_2)\\
=&\int \frac{d^3p}{(2\pi)^3}\ \frac{1}{2p}(1-i\frac{1}{p\tau_1})(1+i\frac{1}{p\tau_2})
\ e^{-ip(\tau_1-\tau_2)+i{\bf p}\cdot({\bf x}_1-{\bf x}_2)} .
\end{split}\end{equation}

If we blindly apply Feynman rules to investigate the effects of the interaction, the
integrations over time give rise to infra-red(IR) divergences at the infinite future
\cite{SH}. 
For example, with ${\lambda}\varphi^3$ interaction we find
\begin{equation}\begin{split}
\int^0_{-\infty}d\tau_1\ \frac{1}{H^4\tau_1^4}\times \{(H\tau)(H\tau_1)
\langle T\varphi(x)\varphi(x_1)\rangle\}^3
\sim\int^0d\tau_1\ \frac{1}{\tau_1^4} .
\end{split}\end{equation}
In the next section, we recall the method to investigate the effects of the interaction which
does not suffer from these IR divergences.

\section{The Schwinger-Keldysh formalism}
\setcounter{equation}{0}

\hspace{0.7cm}
Let us represent the vacuum at $t\to -\infty$ as $|in\rangle$, and $t\to +\infty$ as $|out\rangle$. 
In the Feynman-Dyson formalism, the vacuum expectation value(vev) is essentially given by the transition amplitude 
between $|in\rangle$ and $|out\rangle$
\begin{equation}\begin{split}
\langle \mathcal{O}_H(t)\rangle
&=\langle out| U(+\infty,t)\mathcal{O}_I(t)U(t,-\infty)|in\rangle\\
&=\langle out| T[U(+\infty,-\infty)\mathcal{O}_I(t)]|in\rangle ,
\end{split}\end{equation}
where $\mathcal{O}_H$ and $\mathcal{O}_I$ denote the operators in the Heisenberg and the interaction pictures
respectively.
$U(t_1,t_2)$ is the time translation operator in the interaction picture
\begin{equation}
U(t_1,t_2)=T\left[\exp\left\{-i\int^{t_1}_{t_2}dt\ H_I(t)\right\}\right] .
\end{equation}
It is because $|in\rangle$ is equal to $|out\rangle$ up to a phase due to the time translation invariance.
On the other hand, there is no time translation symmetry in de Sitter space, and so we can't prefix $|out\rangle$.
In this case, we can evaluate the vev only with respect to $|in\rangle$
\begin{equation}\begin{split}
\langle \mathcal{O}_H(t)\rangle
&=\langle in| U(-\infty,t)\mathcal{O}_I(t)U(t,-\infty)|in\rangle\\
&=\langle in| T_C[U\mathcal{O}_I(t)]|in\rangle .
\end{split}\end{equation}
Here $T_C$ denotes the operator ordering specified by the following path
\begin{center}
\includegraphics[width=6cm,clip]{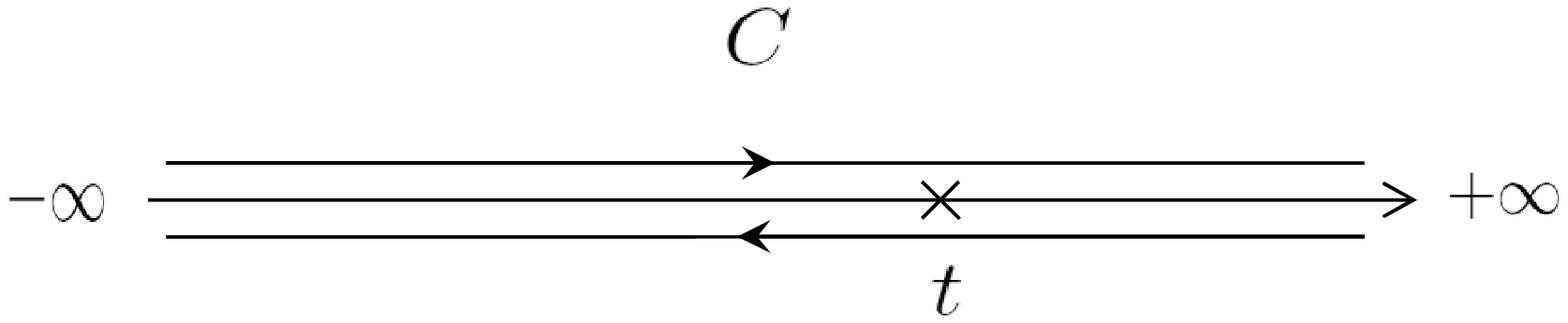}
\end{center}
\begin{equation}
\int_C dt = \int^\infty_{-\infty} dt_+ - \int^\infty_{-\infty} dt_- .
\end{equation}
Because there are two time indices $(+,-)$, 
the propagator has 4 components
\begin{equation}\begin{split}
\Check{G}(x_1,x_2) 
&\equiv \begin{pmatrix} G^{++}(x_1,x_2) & G^{+-}(x_1,x_2) \\ G^{-+}(x_1,x_2) & G^{--}(x_1,x_2) \end{pmatrix} \\
&=\begin{pmatrix} \langle T\varphi(x_1)\varphi(x_2) \rangle & \langle \varphi(x_2)\varphi(x_1) \rangle \\
 \langle \varphi(x_1)\varphi(x_2) \rangle & \langle \Tilde{T}\varphi(x_1)\varphi(x_2) \rangle \end{pmatrix} .
\end{split}\end{equation}
Here $\Tilde{T}$ denotes the antitime-ordering.

In order to investigate the effects of the interaction,
we consider the following Schwinger-Dyson equation for the two point function 
\begin{equation}\begin{split}
\Check{G}(x_1,x_2)
=&\ \Check{G}_0(x_1,x_2)\\ 
&+\int\sqrt{-g_3}d^4x_3\sqrt{-g_4}d^4x_4\ \Check{G}_0(x_1,x_3) 
\begin{pmatrix} 1 & 0 \\ 0 & -1 \end{pmatrix}\\
&\hspace{3.6cm}\times\Check{\Sigma}(x_3,x_4)
\begin{pmatrix} 1 & 0 \\ 0 & -1 \end{pmatrix}
\Check{G}(x_4,x_2) ,
\end{split}\end{equation}
where $G_0$ is the free propagator, $G$ is the full propagator, and $\Sigma$ is the particle's self energy. 
Especially, we focus on the $(-+)$ component of the propagator
\begin{align}
G^{-+}(x_1,x_2)=&\ G_0^{-+}(x_1,x_2)\label{L}\\
&+\int\sqrt{-g_3}d^4x_3\sqrt{-g_4}d^4x_4\ G_0^{-+}(x_1,x_3)\Sigma^{++}(x_3,x_4)G^{++}(x_4,x_2)\notag\\
&-\int\sqrt{-g_3}d^4x_3\sqrt{-g_4}d^4x_4\ G_0^{-+}(x_1,x_3)\Sigma^{+-}(x_3,x_4)G^{-+}(x_4,x_2)\notag\\
&-\int\sqrt{-g_3}d^4x_3\sqrt{-g_4}d^4x_4\ G_0^{--}(x_1,x_3)\Sigma^{-+}(x_3,x_4)G^{++}(x_4,x_2)\notag\\
&+\int\sqrt{-g_3}d^4x_3\sqrt{-g_4}d^4x_4\ G_0^{--}(x_1,x_3)\Sigma^{--}(x_3,x_4)G^{-+}(x_4,x_2)\notag\\
=&\ G^{-+}_0(x_1,x_2)\notag\\
&+\int\sqrt{-g_3}d^4x_3\sqrt{-g_4}d^4x_4\ G_0^{R}(x_1,x_3)\Sigma^{R}(x_3,x_4)G^{-+}(x_4,x_2)\notag\\
&+\int\sqrt{-g_3}d^4x_3\sqrt{-g_4}d^4x_4\ G_0^{R}(x_1,x_3)\Sigma^{-+}(x_3,x_4)G^{A}(x_4,x_2)\notag\\
&+\int\sqrt{-g_3}d^4x_3\sqrt{-g_4}d^4x_4\ G_0^{-+}(x_1,x_3)\Sigma^{A}(x_3,x_4)G^{A}(x_4,x_2).\notag
\end{align}
Here we have introduced the retarded and the advanced propagators as follows
\begin{equation}\begin{split}
&G^R(x_1,x_2)\equiv \theta(t_1-t_2)[G^{-+}(x_1,x_2)-G^{+-}(x_1,x_2)] ,\\
&G^A(x_1,x_2)\equiv -\theta(t_2-t_1)[G^{-+}(x_1,x_2)-G^{+-}(x_1,x_2)] .
\end{split}\end{equation}

We observe that the integrations over time are bounded by $t_1$ or $t_2$ because of the causality.
In the same way, the following identity also holds
\begin{equation}\begin{split}
G^{-+}(x_1,x_2)
=&\ G^{-+}_0(x_1,x_2)\\
&+\int\sqrt{-g_3}d^4x_3\sqrt{-g_4}d^4x_4\ G^{R}(x_1,x_3)\Sigma^{R}(x_3,x_4)G^{-+}_0(x_4,x_2)\\
&+\int\sqrt{-g_3}d^4x_3\sqrt{-g_4}d^4x_4\ G^{R}(x_1,x_3)\Sigma^{-+}(x_3,x_4)G^{A}_0(x_4,x_2)\\
&+\int\sqrt{-g_3}d^4x_3\sqrt{-g_4}d^4x_4\ G^{-+}(x_1,x_3)\Sigma^{A}(x_3,x_4)G^{A}_0(x_4,x_2) .
\end{split}\label{R}\end{equation}
In this formalism, the integrations over time are manifestly finite due to the causality.
This formalism is called Schwinger-Keldysh formalism. 
In order to understand the effects of the interaction, 
we derive a Boltzmann equation on the dS background from a Schwinger-Dyson equation in the next section. 

\section{Boltzmann equations from Schwinger-Dyson equations}
\setcounter{equation}{0}

\hspace{0.7cm}
In a time dependent background, we need to consider excited states in general. 
For such a state, the expectation value of the number operator $\langle a^\dagger a \rangle$ is non-vanishing.
We introduce a distribution function $f$ for scalar particles as follows
\begin{equation}
\langle a^\dagger_{{\bf p}} a_{{\bf q}} \rangle \equiv f({\bf p})\times (2\pi)^3\delta^{(3)}
({\bf p}-{\bf q}) .
\label{number}\end{equation}
One of our main objectives in this paper is to understand the time dependence of the distribution function 
$f$ due to the interaction. 
We utilize a Boltzmann equation for this purpose. 
Boltzmann equations govern the time evolution of the distribution functions.
They are widely used to study non-equilibrium physics. 
In fact there is a long history of the microscopic derivation 
of Boltzmann equations in non-equilibrium physics using Schwinger-Keldysh formalism
\cite{KB}\cite{Kd}\cite{kita}.  
In this paper, we systematically investigate the propagator in dS space
from a Schwinger-Dyson equation.

We assume that the full propagator in de Sitter space has the following form
\begin{equation}\begin{split}
G^{-+}(x_1,x_2)=\ \ &\int \frac{d^3p}{(2\pi)^3}\ 
\big[(1+f(p,\tau_c))Z(p,\tau_c)\phi_{{\bf p}}(x_1)\phi_{{\bf p}}^*(x_2)\\
&\hspace{2.4cm}+f(p,\tau_c)Z^*(p,\tau_c)\phi_{{\bf p}}^*(x_1)\phi_{{\bf p}}(x_2)\big]\\
+&\int_{\varepsilon>0} \frac{d\varepsilon d^3p}{(2\pi)^4}
\ \frac{1}{2\varepsilon}[F_+(\varepsilon,p,\tau_c)\ e^{-i\varepsilon(\tau_1-\tau_2)+i{\bf p}\cdot
({\bf x}_1-{\bf x}_2)}\\
&\hspace{2.8cm}+F_-(\varepsilon,p,\tau_c)\ e^{+i\varepsilon(\tau_1-\tau_2)-i{\bf p}\cdot({\bf x}_1-{\bf x}_2)}] .
\end{split}\label{full}\end{equation}
The propagator depends on the average and the relative time:
\begin{equation}\begin{split}
\tau_c\equiv \frac{\tau_1+\tau_2}{2},\hspace{0.8cm}\bar{\tau}\equiv \tau_1-\tau_2 .
\end{split}\end{equation}
It consists of the on-shell part and the off-shell part.
In the on-shell part, we have introduced the wave function renormalization factor
$Z(p,\tau_c)$.
The off-shell part depends on the spectral function $F_{\pm}(\varepsilon,p,\tau_c)$.
We assume that $f, Z, F_{\pm}$ evolve with the average time $\tau_c$. 
We investigate the propagator in the region:
\begin{equation}
|\tau_c|\gg |\bar{\tau}|,\hspace{0.8cm}|\tau_c|\gg 1/p .
\end{equation}
The second assumption implies that we investigate the propagator well inside the cosmological horizon.

From (\ref{L}) and (\ref{R}), we can derive the following identity
\begin{equation}\begin{split}
&G_0^{-1}|_1G^{-+}(x_1,x_2)-G_0^{-1}|_2G^{-+}(x_1,x_2)\\
=&+\sqrt{-g_1}\int\sqrt{-g_3}d^4x_3\ \Sigma^R(x_1,x_3)G^{-+}(x_3,x_2)\\
&+\sqrt{-g_1}\int\sqrt{-g_3}d^4x_3\ \Sigma^{-+}(x_1,x_3)G^A(x_3,x_2)\\
&-\sqrt{-g_2}\int\sqrt{-g_3}d^4x_3\ G^R(x_1,x_3)\Sigma^{-+}(x_3,x_2)\\
&-\sqrt{-g_2}\int\sqrt{-g_3}d^4x_3\ G^{-+}(x_1,x_3)\Sigma^{A}(x_3,x_2) .
\end{split}\label{Boltzmann}\end{equation}
By putting the expression for the full propagator (\ref{full}) into the left-hand side of 
the Schwinger-Dyson equation (\ref{Boltzmann}), we obtain
\begin{equation}\begin{split}
&\ \ G_0^{-1}|_1G^{-+}(x_1,x_2)-G_0^{-1}|_2G^{-+}(x_1,x_2)\\
\sim&\ \int \frac{d^3p}{(2\pi)^3}\ \Big[
\big(\frac{\partial}{\partial \tau_c}
+\frac{i}{p}\frac{\partial^2}{\partial\bar{\tau}\partial\tau_c}\big)
\{(1+f(p,\tau_c))Z(p,\tau_c)\}\times
e^{-ip\bar{\tau}+i{\bf p}\cdot\bar{{\bf x}}}\\
&\hspace{2.8cm}-\big(\frac{\partial }{\partial \tau_c}
-\frac{i}{p}\frac{\partial^2}{\partial\bar{\tau}\partial\tau_c}\big)
\{f(p,\tau_c)Z^*(p,\tau_c)\}\times
e^{+ip\bar{\tau}-i{\bf p}\cdot\bar{{\bf x}}}\Big]\\
&+\int_{\varepsilon>0} \frac{d\varepsilon d^3p}{(2\pi)^4}
\ \Big[\big(\frac{\partial}{\partial \tau_c}
+\frac{i}{\varepsilon}\frac{\partial^2}{\partial\bar{\tau}\partial\tau_c}\big)F_+(\varepsilon,p,\tau_c)\times
e^{-i\varepsilon\bar{\tau}+i{\bf p}\cdot\bar{{\bf x}}}\\
&\hspace{3.6cm}-\big(\frac{\partial}{\partial \tau_c}
-\frac{i}{\varepsilon}\frac{\partial^2}{\partial\bar{\tau}\partial\tau_c}\big)F_-(\varepsilon,p,\tau_c)\times
e^{+i\varepsilon\bar{\tau}-i{\bf p}\cdot\bar{{\bf x}}}\Big] .
\end{split}\label{derivative}\end{equation}
Here we recall the following definitions
\begin{equation}\begin{split}
\hspace{1.4cm}&G_0^{-1} \equiv i(\partial_\tau^2-\partial_{{\bf x}}^2-\frac{2}{\tau^2}),\\
&\hspace{0.8cm}G_0^{-1}|_1G^R(x_1,x_2)=\delta^{(4)}(x_1-x_2),\\
&\hspace{0.8cm}G_0^{-1}|_2G^A(x_1,x_2)=\delta^{(4)}(x_1-x_2) .\\
\end{split}\end{equation}
In (\ref{derivative}) we have shown the leading terms in the power series expansion of $1/p\tau_c$.

The right-hand side of Eq.(\ref{Boltzmann}) corresponds to the collision term $C[f]$.
In this paper, we investigate the effects of the interaction in ${\lambda}\varphi^3$ theory 
at the one loop level. 
We subsequently find that this theory captures the essential features of more generic
field theories such as $g\varphi^4$ theory.
The self-energy is
\begin{center}
\includegraphics[width=3.5cm,clip]{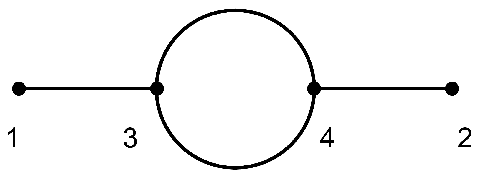}
\end{center}
\begin{equation}
\Sigma^{ij}(x_3,x_4)=\frac{(-i\lambda)^2}{2}G^{ij}(x_3,x_4)G^{ij}(x_3,x_4)
,\hspace{0.4cm}i,j=+,- .
\label{Sigma}\end{equation}
To the leading order in perturbation theory, 
we can approximate that $f(p,\tau_c)=f(p)$, $Z(p,\tau_c)=1$, $F_\pm(\varepsilon,p,\tau_c)=0$ in the collision term. 
We also expand the collision term by the power series in $1/|p\tau_c|$ type factors which can be justified well inside
the cosmological horizon.
It is a kind of the derivative expansion of the Moyal product in the Wigner representation.  
We indeed find the particle production
effects due to the non-conservation of the energy in this expansion. 

In this investigation, we need to perform the following integrations at the interaction vertices.
\begin{equation}
\int^{\tau_i}_{-\infty}d\tau_3\ \frac{1}{\tau_3^n}e^{i(\varepsilon\pm p)\tau_3}
\hspace{0.8cm}n\in {\bf N},\hspace{0.4cm}i=1,2 ,
\end{equation}
where $\varepsilon =\pm p_1\pm p_2$. 
We evaluate these integrations in the assumption $|(\varepsilon\pm p)\tau_i|\gg 1$ .
For our purpose, it suffices to evaluate them to the next leading order
\begin{equation}
\int^{\tau_i}_{-\infty}d\tau_3\ \frac{1}{\tau_3^n}e^{i(\varepsilon\pm p)\tau_3}
\sim \ e^{i(\varepsilon\pm p)\tau_i}\times
\left[\frac{1}{i(\varepsilon\pm p)\tau_i^n}+\frac{-n}{(\varepsilon\pm p)^2\tau_i^{n+1}}\right] .
\label{Gamma}\end{equation}
By using these approximations, we derive a Boltzmann equation in de Sitter space. 
In what follows, we investigate the collision terms and their properties in detail.

We henceforth suppress the following integration factor in the propagator 
\begin{equation}
\int\frac{d^3p}{(2\pi)^3}e^{i{\bf p}\cdot\bar{{\bf x}}} .
\end{equation}
In other words we work in the momentum space by performing the Fourier transformation with respect to 
the spacial coordinate $\bar{{\bf x}}$.

\newpage

\subsection{The structure of the collision term}

\hspace{0.7cm}
From the Schwinger-Dyson equation (\ref{Boltzmann}), we observe that 
the collision term has the on-shell part and the off-sell part. 
Firstly, the on-shell part comes from the following contributions 
\begin{equation}\begin{split}
C_{on}[f]=
&+\sqrt{-g_1}\int\sqrt{-g_3}d^4x_3\ \Sigma^R(x_1,x_3)G^{-+}(x_3,x_2)\\
&-\sqrt{-g_2}\int\sqrt{-g_3}d^4x_3\ G^{-+}(x_1,x_3)\Sigma^{A}(x_3,x_2)\\
\propto &\ e^{\mp ip\bar{\tau}} .
\end{split}\label{origin1}\end{equation}
We evaluate the on-shell part  
to the leading non-trivial order $\mathcal{O}(1/\tau_c^3)$ as
\begin{align}
&C_{on}[f]\notag\\
=&-(1+f(p))e^{-ip\bar{\tau}}\frac{\lambda^2}{16\pi p^2H^2}\times\label{on}\\
&\hspace{0.8cm}\Big[\int^\infty_p\frac{d\varepsilon}{2\pi}
\big\{(\frac{1}{\varepsilon-p}+\frac{1}{\varepsilon+p})\frac{i\bar{\tau}}{\tau_c^3}
+(\frac{1}{(\varepsilon-p)^2}-\frac{1}{(\varepsilon+p)^2})\frac{-1}{\tau_c^3}\big\}\notag\\
&\hspace{1.6cm}\times\int^{\frac{\varepsilon+p}{2}}_{\frac{\varepsilon-p}{2}}dp_1
\big\{(1+f(p_1))(1+f(\varepsilon-p_1))-f(p_1)f(\varepsilon-p_1)\big\}\notag\\
&\hspace{0.8cm}+2\int^p_0\frac{d\varepsilon}{2\pi}
\big\{(\frac{1}{\varepsilon-p}+\frac{1}{\varepsilon+p})\frac{i\bar{\tau}}{\tau_c^3}
+(\frac{1}{(\varepsilon-p)^2}-\frac{1}{(\varepsilon+p)^2})\frac{-1}{\tau_c^3}\big\}\notag\\
&\hspace{1.6cm}\times\int^\infty_{\frac{\varepsilon+p}{2}}dp_1\big\{(1+f(p_1))f(p_1-\varepsilon)-f(p_1)
(1+f(p_1-\varepsilon))\big\}\hspace{0.4cm}\Big]\notag\\
&+f(p)\ e^{+ip\bar{\tau}}\frac{\lambda^2}{16\pi p^2H^2}\times\notag\\
&\hspace{0.8cm}\Big[\int^\infty_p\frac{d\varepsilon}{2\pi}
\big\{(\frac{1}{\varepsilon-p}+\frac{1}{\varepsilon+p})\frac{-i\bar{\tau}}{\tau_c^3}
+(\frac{1}{(\varepsilon-p)^2}-\frac{1}{(\varepsilon+p)^2})\frac{-1}{\tau_c^3}\big\}\notag\\
&\hspace{1.6cm}\times\int^{\frac{\varepsilon+p}{2}}_{\frac{\varepsilon-p}{2}}dp_1
\big\{(1+f(p_1))(1+f(\varepsilon-p_1))-f(p_1)f(\varepsilon-p_1)\big\}\notag\\
&\hspace{0.8cm}+2\int^p_0\frac{d\varepsilon}{2\pi}
\big\{(\frac{1}{\varepsilon-p}+\frac{1}{\varepsilon+p})\frac{-i\bar{\tau}}{\tau_c^3}
+(\frac{1}{(\varepsilon-p)^2}-\frac{1}{(\varepsilon+p)^2})\frac{-1}{\tau_c^3}\big\}\notag\\
&\hspace{1.6cm}\times\int^\infty_{\frac{\varepsilon+p}{2}}dp_1\big\{(1+f(p_1))f(p_1-\varepsilon)-f(p_1)
(1+f(p_1-\varepsilon))\big\}\hspace{0.4cm}\Big].\notag
\end{align}
See Appendix A for the details of the calculation.

Secondly, the off-shell part originates from the following contribution
\begin{equation}\begin{split}
C_{off}[f]=
&+\sqrt{-g_1}\int\sqrt{-g_3}d^4x_3\ \Sigma^{-+}(x_1,x_3)G^A(x_3,x_2)\\
&-\sqrt{-g_2}\int\sqrt{-g_3}d^4x_3\ G^R(x_1,x_3)\Sigma^{-+}(x_3,x_2)\\
\propto &\ \int^\infty_0dp_1\int^{p_1+p}_{|p_1-p|}dp_2\ e^{-i(\pm p_1\pm p_2)\bar{\tau}} . 
\end{split}\end{equation}
The off-shell part is also calculated to $\mathcal{O}(1/\tau_c^3)$ as
\begin{align}
&C_{off}[f]\notag\\
=&+\frac{\lambda^2}{16\pi p^2H^2}\times\label{off}\\
&\hspace{0.4cm}\Big[\int^\infty_p\frac{d\varepsilon}{2\pi}\ e^{-i\varepsilon\bar{\tau}}
(\frac{1}{(\varepsilon-p)^2}-\frac{1}{(\varepsilon+p)^2})\frac{-1}{\tau_c^3}
\int^{\frac{\varepsilon+p}{2}}_{\frac{\varepsilon-p}{2}}dp_1\ (1+f(p_1))(1+f(\varepsilon-p_1))\notag\\
&\hspace{0.8cm}+2\int^p_0\frac{d\varepsilon}{2\pi}\ e^{-i\varepsilon\bar{\tau}}
(\frac{1}{(\varepsilon-p)^2}-\frac{1}{(\varepsilon+p)^2})\frac{-1}{\tau_c^3}
\int^\infty_{\frac{\varepsilon+p}{2}}dp_1\ (1+f(p_1))f(p_1-\varepsilon)\hspace{0.4cm}\Big]\notag\\
&-\frac{\lambda^2}{16\pi p^2H^2}\times\notag\\
&\hspace{0.4cm}\Big[\int^\infty_p\frac{d\varepsilon}{2\pi}\ e^{+i\varepsilon\bar{\tau}}
(\frac{1}{(\varepsilon-p)^2}-\frac{1}{(\varepsilon+p)^2})\frac{-1}{\tau_c^3}
\int^{\frac{\varepsilon+p}{2}}_{\frac{\varepsilon-p}{2}}dp_1\ f(p_1)f(\varepsilon-p_1)\notag\\
&\hspace{0.8cm}+2\int^p_0\frac{d\varepsilon}{2\pi}\ e^{+i\varepsilon\bar{\tau}}
(\frac{1}{(\varepsilon-p)^2}-\frac{1}{(\varepsilon+p)^2})\frac{-1}{\tau_c^3}
\int^\infty_{\frac{\varepsilon+p}{2}}dp_1\ f(p_1)(1+f(p_1-\varepsilon))\ \Big].\notag
\end{align}

We note that the both on-shell (\ref{on}) and off-shell (\ref{off}) collision terms have infra-red divergences at 
$\varepsilon=p$. There is a standard procedure to deal with this problem in massless field theory and we find
that it also works here. First of all, we need to recall that any experiment has a finite energy resolution
$\Delta\varepsilon$. So we need to add the on-shell and off-shell collision terms
within the energy resolution $\Delta\varepsilon$.
We first divide the integration range of $C_{off}[f]$ as follows 
\begin{equation}\begin{split}
\int_p^\infty=
\int_{p+\Delta\varepsilon}^\infty
+\int_p^{p+\Delta\varepsilon}\ ,\hspace{0.4cm}
\int_0^p=
\int^{p-\Delta\varepsilon}_0
+\int^p_{p-\Delta\varepsilon} .
\end{split}\label{redef}\end{equation}
We then redefine the on-shell term $C_{on}'[f]$ and the off-shell term $C_{off}'[f]$
by transferring the contribution of $C_{off}[f]$ within the energy resolution
$p-\Delta\varepsilon\leq\varepsilon\leq p+\Delta\varepsilon$ to 
$C_{on}[f]$. 
The explicit expressions are shown in Appendix A.

When $f(p)=0$, we find that infra-red divergences cancel out in this procedure.
In the next subsection, we investigate the case when $f$ is a thermal distribution.
For a generic distribution, the cancellation does not take place and we seem to face linear IR divergences.
However there is no real infra-red divergence in our problem since the time integration range in (\ref{Gamma}) 
is bounded by $\tau_c$. We thus argue that the linear divergence should be cut-off at $|p-\varepsilon| \sim 1/|\tau_c|$.

Before investigating the thermal distribution case, 
we point out the difference between Minkowski space and dS space with respect to the collision term. 
In Minkowski space, the collision term does not have the off-shell term due to the time translation symmetry
\begin{equation}\begin{split}
C_{off}[f]&\propto \int\frac{d\varepsilon}{2\pi}\ 2\pi\delta(\varepsilon-p)e^{\mp i\varepsilon\bar{\tau}}=
\ e^{\mp ip\bar{\tau}}\ 
\Longrightarrow \ C_{off}'[f]=0 .
\end{split}\end{equation}
On the other hand, as we observe in (\ref{off}), the collision term in dS space has the off-shell term 
due to the absence of the time translation symmetry. 
This is why we have introduced the spectral function $F_{\pm}(\varepsilon,p,\tau_c)$ in the full propagator (\ref{full}). 
 
\subsection{Thermal distribution case}

\hspace{0.7cm}
We focus on the case that the initial distribution function is thermal in this subsection
\begin{equation}
f(p)=\frac{1}{e^{\beta p}-1} ,
\end{equation}
where $\beta$ is an inverse temperature. 
In Minkowski space
the thermal distribution is obtained as the solution of the Boltzmann equation.
On the other hand, we find that the collision term in dS space is non-vanishing even for the thermal distribution. 

The off-shell collision term can be evaluated as follows
\begin{align}
&C_{off}'[f]\notag\\
=&+\frac{\lambda^2}{16\pi pH^2}\times\label{off_2}\\
&\hspace{0.4cm}\Big[\int^\infty_{p+\Delta\varepsilon}\frac{d\varepsilon}{2\pi}\ (1+f(\varepsilon))e^{-i\varepsilon\bar{\tau}}
(\frac{1}{(\varepsilon-p)^2}-\frac{1}{(\varepsilon+p)^2})\frac{-1}{\tau_c^3}
(1+ G(\varepsilon, p, \beta )
)\notag\\
&\hspace{0.8cm}+\int^{p-\Delta\varepsilon}_0\frac{d\varepsilon}{2\pi}\ (1+f(\varepsilon))e^{-i\varepsilon\bar{\tau}}
(\frac{1}{(\varepsilon-p)^2}-\frac{1}{(\varepsilon+p)^2})\frac{-1}{\tau_c^3}
G(\varepsilon, p, \beta )
\hspace{0.4cm}\Big]\notag\\
&-\frac{\lambda^2}{16\pi pH^2}\times\notag\\
&\hspace{0.4cm}\Big[\int^\infty_{p+\Delta\varepsilon}\frac{d\varepsilon}{2\pi}\ f(\varepsilon)e^{+i\varepsilon\bar{\tau}}
(\frac{1}{(\varepsilon-p)^2}-\frac{1}{(\varepsilon+p)^2})\frac{-1}{\tau_c^3}
(1+G(\varepsilon, p, \beta )
)\notag\\
&\hspace{0.8cm}+\int^{p-\Delta\varepsilon}_0\frac{d\varepsilon}{2\pi}\ f(\varepsilon)e^{+i\varepsilon\bar{\tau}}
(\frac{1}{(\varepsilon-p)^2}-\frac{1}{(\varepsilon+p)^2})\frac{-1}{\tau_c^3}
G(\varepsilon, p, \beta )
\hspace{0.4cm}\Big],\notag
\end{align}
where 
\begin{equation}
G(\varepsilon, p, \beta )\equiv
\frac{2}{\beta p}\log\left(\frac{1-e^{-\beta\frac{\varepsilon+p}{2}}}{1-e^{-\beta\frac{|\varepsilon-p|}{2}}}\right) .
\label{Gfn}\end{equation}
We note that the above expression is of the following form
\begin{equation}\begin{split}
C_{off}'[f]=\int_{\varepsilon > 0}\frac{d\varepsilon}{2\pi}
\left((1+f(\varepsilon))A(\varepsilon,p,\tau_c)e^{-i\varepsilon\bar{\tau}}
-f(\varepsilon)A^*(\varepsilon,p,\tau_c)e^{i\varepsilon\bar{\tau}}\right) .
\end{split}\label{Z_off}\end{equation}
It is consistent with our ansatz for the full propagator (\ref{full}).

Finally the on-shell collision term is evaluated as follows
\begin{align}
&C_{on}'[f]\notag\\
=&-\frac{\lambda^2}{16\pi pH^2}(1+f(p))e^{-ip\bar{\tau}}\times\label{on_3}\\
&\hspace{0.4cm}\Big[\int^\infty_{p+\Delta\varepsilon}
\frac{d\varepsilon}{2\pi}
\big\{(\frac{1}{\varepsilon-p}+\frac{1}{\varepsilon+p})\frac{i\bar{\tau}}{\tau_c^3}
+(\frac{1}{(\varepsilon-p)^2}-\frac{1}{(\varepsilon+p)^2})\frac{-1}{\tau_c^3}\big\}
(1+G(\varepsilon, p, \beta )
)\notag\\
&\hspace{0.8cm}+\int^{p-\Delta\varepsilon}_0
\frac{d\varepsilon}{2\pi}
\big\{(\frac{1}{\varepsilon-p}+\frac{1}{\varepsilon+p})\frac{i\bar{\tau}}{\tau_c^3}
+(\frac{1}{(\varepsilon-p)^2}-\frac{1}{(\varepsilon+p)^2})\frac{-1}{\tau_c^3}\big\}
G(\varepsilon, p, \beta )
\ \Big]\notag\\
&+\frac{\lambda^2}{16\pi pH^2}\ f(p)e^{+ip\bar{\tau}}\times\notag\\
&\hspace{0.4cm}\Big[\int^\infty_{p+\Delta\varepsilon}
\frac{d\varepsilon}{2\pi}
\big\{(\frac{1}{\varepsilon-p}+\frac{1}{\varepsilon+p})\frac{-i\bar{\tau}}{\tau_c^3}
+(\frac{1}{(\varepsilon-p)^2}-\frac{1}{(\varepsilon+p)^2})\frac{-1}{\tau_c^3}\big\}
(1+G(\varepsilon, p, \beta )
)\notag\\
&\hspace{0.8cm}+\int^{p-\Delta\varepsilon}_0
\frac{d\varepsilon}{2\pi}
\big\{(\frac{1}{\varepsilon-p}+\frac{1}{\varepsilon+p})\frac{-i\bar{\tau}}{\tau_c^3}
+(\frac{1}{(\varepsilon-p)^2}-\frac{1}{(\varepsilon+p)^2})\frac{-1}{\tau_c^3}\big\}
G(\varepsilon, p, \beta )
\ \Big]\notag\\
&\notag\\
&+\frac{\lambda^2}{32\pi^2pH^2}\frac{-1}{\tau_c^3}\log|\Delta\varepsilon\tau_c|\ f'(p)e^{-ip\bar{\tau}}
-\frac{\lambda^2}{32\pi^2pH^2}\frac{-1}{\tau_c^3}\log|\Delta\varepsilon\tau_c|\ f'(p)e^{+ip\bar{\tau}}.\notag
\end{align}
The details of its derivation can be found in Appendix A.

Here we have cut-off the IR log divergences when $|\varepsilon -p | \sim 1/|\tau_c|$
because our time integration (\ref{Gamma}) does not diverge even when $\varepsilon=p$.
From the on-shell collision term (\ref{on_3}), we observe that it is necessary
to introduce the wave function renormalization factor $Z(p,\tau_c)$.
In the last line, we find that the remaining logarithmic IR contribution leads to the modification of 
the thermal distribution function $\delta f(p,\tau_c)$.

So far, we have focused on the IR singularities due to the interaction. 
Of course, there are also the ultra-violet(UV) divergences in the collision term. 
The off-shell part (\ref{off}) does not have the UV divergences because of the exponentially oscillating factor. 
We also assume that a
generic distribution function vanishes exponentially at the UV region like the Bose distribution
\begin{equation}
f(p_i)\approx \frac{1}{e^{\beta p_i}-1}\ \to\ 0 .
\end{equation}
From these facts, the UV divergences in the collision term is estimated as follows
\begin{equation}\begin{split}
C[f]_{UV}=&\ C'_{on}[f]\\
\approx&-\frac{\lambda^2}{16\pi pH^2}(1+f(p))e^{-ip\bar{\tau}}
\int^{\Lambda_{UV} e^{Ht}}_{p+\Delta\varepsilon}
\frac{d\varepsilon}{2\pi}\ 
(\frac{1}{\varepsilon-p}+\frac{1}{\varepsilon+p})\frac{i\bar{\tau}}{\tau_c^3}\\
&+\frac{\lambda^2}{16\pi pH^2}\ f(p)e^{+ip\bar{\tau}}
\int^{\Lambda_{UV} e^{Ht}}_{p+\Delta\varepsilon}
\frac{d\varepsilon}{2\pi}\ 
(\frac{1}{\varepsilon-p}+\frac{1}{\varepsilon+p})\frac{-i\bar{\tau}}{\tau_c^3}\\
=&-i\frac{\lambda^2}{16\pi^2}\frac{2\bar{\tau}}{H^2\tau_c^3}\log\frac{\Lambda_{UV}e^{Ht}}{q}
\times (1+f(p))\ \frac{1}{2p}e^{-ip\bar{\tau}}\\
& +i\frac{\lambda^2}{16\pi^2}\frac{2\bar{\tau}}{H^2\tau_c^3}\log\frac{\Lambda_{UV}e^{Ht}}{q}
\times f(p)\ \frac{1}{2p}e^{+ip\bar{\tau}} .
\end{split}\label{UVexp}\end{equation}
Since the integral is logarithmically divergent, we need to introduce a UV cut-off.
We argue that we need to cut-off the integral at a fixed physical energy scale $\Lambda_{UV}$.
As the physical energy is $\varepsilon H |\tau|$, this prescription leads to a time dependent UV cut-off
$\Lambda_{UV} /H|\tau| = \Lambda_{UV}e^{Ht}$ in the above expression.
We believe that this is a physically sensible prescription which is consistent with general covariance.
In this prescription, the degrees of freedom inside the cosmological horizon remain the same with respect 
to time.
The IR cut-off is provided by our energy resolution $\Delta\varepsilon$ in (\ref{UVexp})
as the IR singularity is canceled by 
the off-shell contribution. The final expression logarithmically depends on the virtuality
$q^2\equiv(p+\Delta\varepsilon)^2-p^2$.

This UV divergence is renormalized by introducing a mass counter term 
in the action which leads to the following collision term
\begin{equation}\begin{split}
C[f]_{\delta m^2}=&+i\frac{2\bar{\tau}}{H^2\tau_c^3}\delta m^2\times(1+f(p))\frac{1}{2p}e^{-ip\bar{\tau}}\\
&-i\frac{2\bar{\tau}}{H^2\tau_c^3}\delta m^2\times f(p)\frac{1}{2p}e^{+ip\bar{\tau}} ,\\
\delta m^2=&\frac{\lambda^2}{16\pi^2}\log\frac{\Lambda_{UV}e^{Ht}}{\mu} ,
\end{split}\end{equation}
where $\mu$ is the renormalization scale.
After the renormalization, we obtain the following effective mass
\begin{equation}\begin{split}
m_{eff}^2=\frac{\lambda^2}{16\pi^2}\left(\log \frac{q}{\mu}
-\frac{1}{\beta p}
\int_0^\infty d\varepsilon \big\{\frac{1}{\varepsilon -p}+\frac{1}{\varepsilon +p}\big\}
\log\left(\frac{1-e^{-\beta (\varepsilon +p)/2}}{1-e^{-\beta |\varepsilon -p|/2}}\right)\right) ,
\end{split}\end{equation}
including the finite temperature correction. In the zero temperature limit,
it agrees with the renormalized mass in the flat space.

The IR logarithm in the collision term (\ref{on_3}) leads to the change of the distribution function 
as we solve the Boltzmann equation
\begin{equation}\begin{split}
\delta f(p,\tau_c)&=\frac{\lambda^2}{64\pi^2p}\frac{1}{H^2\tau_c^2}\log|\Delta\varepsilon\tau_c|f'(p)\\
&=-\frac{\lambda^2}{64\pi^2p}\frac{1}{H^2\tau_c^2}\log|\Delta\varepsilon\tau_c|
\frac{\beta}{e^{\beta p}-1}\frac{e^{\beta p}}{e^{\beta p}-1} .
\end{split}\label{f}\end{equation}
The wave function renormalization factor is determined as
\begin{equation}\begin{split}
&\ \delta Z(p,\tau_c)\\
=&-\frac{\lambda^2}{32\pi pH^2}\times\\
&\hspace{0.4cm}\Big[
\int^\infty_{p+\Delta\varepsilon}\frac{d\varepsilon}{2\pi}
(\frac{1}{(\varepsilon-p)^2}-\frac{1}{(\varepsilon+p)^2})\frac{1}{\tau_c^2}
(1+G(\varepsilon, p, \beta ))
\\
&\ \hspace{3.6cm}+\int^{p-\Delta\varepsilon}_0\frac{d\varepsilon}{2\pi}
(\frac{1}{(\varepsilon-p)^2}-\frac{1}{(\varepsilon+p)^2})\frac{1}{\tau_c^2}
G(\varepsilon, p, \beta )
\ \Big] .
\end{split}\label{Z}\end{equation}
The off-shell part of the propagator is determined in terms $F_{\pm}$ as
\begin{equation}\begin{split}
&F_{+}(\varepsilon,p,\tau_c)\\
=&+\frac{\lambda^2}{32\pi pH^2}(1+f(\varepsilon))\times\\
&\hspace{0.4cm}\Big[\ \theta(\varepsilon-p)
(\frac{1}{(\varepsilon-p)^2}-\frac{1}{(\varepsilon+p)^2})\frac{1}{\tau_c^2}
(1+G(\varepsilon, p, \beta )
)\\
&\hspace{0.4cm}+\theta(p-\varepsilon)
(\frac{1}{(\varepsilon-p)^2}-\frac{1}{(\varepsilon+p)^2})\frac{1}{\tau_c^2}
G(\varepsilon, p, \beta )
\hspace{0.8cm}\Big] ,
\end{split}\label{F+}\end{equation}
\begin{equation}\begin{split}
&F_{-}(\varepsilon,p,\tau_c)\\
=&+\frac{\lambda^2}{32\pi pH^2}f(\varepsilon)\times\\
&\hspace{0.4cm}\Big[\ \theta(\varepsilon-p)
(\frac{1}{(\varepsilon-p)^2}-\frac{1}{(\varepsilon+p)^2})\frac{1}{\tau_c^2}
(1+G(\varepsilon, p, \beta )
)\\
&\hspace{0.4cm}+\theta(p-\varepsilon)
(\frac{1}{(\varepsilon-p)^2}-\frac{1}{(\varepsilon+p)^2})\frac{1}{\tau_c^2}
G(\varepsilon, p, \beta )
\hspace{0.8cm}\Big] .
\end{split}\label{F-}\end{equation}
We observe that the on-shell weight represented by the wave function renormalization factor $Z$
is reduced from the unity in a consistent way with the off-shell spectral weight. 
In this sense unitarity is respected by the interaction.

We have thus determined the full propagator inside the cosmological horizon
to the leading order of the perturbation theory.
We have found that the full propagator which is characterized by (\ref{f}), (\ref{Z}), (\ref{F+}), (\ref{F-}) 
depends on $\tau_c$.  At first sight, it appears to change with cosmic evolution.
More and more off-shell states are created with a lapse of time as on-shell states are correspondingly reduced.
However we may represent (\ref{f}), (\ref{Z}), (\ref{F+}), (\ref{F-}) by the physical quantities, 
\begin{equation}\begin{split}
X\equiv {x\over H|\tau|},\hspace{0.4cm}
P\equiv H|\tau|p,\hspace{0.4cm}
\Delta E\equiv H|\tau|\Delta\varepsilon,\hspace{0.4cm}
T \equiv H|\tau|\frac{1}{\beta},\hspace{0.4cm}
M \equiv H|\tau|\mu .
\end{split}\end{equation}
In terms of the physical quantities, the full propagator of the original scalar field
at the equal time $\bar{\tau}=0$ is
\begin{equation}\begin{split}
G^{-+}(x_1,x_2)=&\int\frac{d^3P}{(2\pi)^32P}\ \big(1+2(f+\delta f)\big)(1+\delta Z)
\Big\{1+(1-\frac{m^2_{eff}}{2H^2})\frac{H^2}{P^2}\Big\}e^{iP\cdot \bar{X}}\\
&+\int_{E>0}\frac{dEd^3P}{(2\pi)^42E}\ (F_++F_-)e^{iP\cdot \bar{X}},
\end{split}\end{equation}
\begin{equation}\begin{split}
m^2_{eff}
=\frac{\lambda^2}{32\pi^2}\left(\log \frac{Q^2}{M^2} 
-\frac{2T}{P}
\int_0^\infty dE \big\{\frac{1}{E -P}+\frac{1}{E +P}\big\}
\log\left(\frac{1-e^{-(E +P)/2T}}{1-e^{- |E -P|/2T}}\right)\right),
\end{split}\end{equation}
\begin{equation}\begin{split}
\delta f=\frac{\lambda^2}{64\pi^2P}\frac{\partial f(P,T)}{\partial P}\log\frac{\Delta E}{H} ,
\end{split}\label{mdf}\end{equation}
\begin{equation}\begin{split}
\delta Z
=&-\frac{\lambda^2}{32\pi P}\times\\
&\hspace{0.4cm}\Big[\ \frac{1}{2\pi}(\frac{1}{\Delta E}-\frac{1}{2 P})+\int^\infty_{P+\Delta E}\frac{dE}{2\pi}
(\frac{1}{(E-P)^2}-\frac{1}{(E+P)^2})
\frac{2T}{P}\log\left(\frac{1-e^{-\frac{E+P}{2T}}}{1-e^{-\frac{E-P}{2T}}}\right)\\
&\hspace{2.8cm}+\int^{P-\Delta E}_0\frac{dE}{2\pi}
(\frac{1}{(E-P)^2}-\frac{1}{(E+P)^2})
\frac{2T}{P}\log\left(\frac{1-e^{-\frac{E+P}{2T}}}{1-e^{-\frac{P-E}{2T}}}\right)
\Big] ,
\end{split}\end{equation}
\begin{equation}\begin{split}
F_{+}
=&\frac{\lambda^2}{32\pi P}(1+f(E,T))\times\\
&\hspace{0.4cm}\Big[\ \theta(E-P)
(\frac{1}{(E-P)^2}-\frac{1}{(E+P)^2})
(1+\frac{2T}{P}\log\left(\frac{1-e^{-\frac{E+P}{2T}}}{1-e^{-\frac{E-P}{2T}}}\right))\\
&\hspace{0.8cm}+\theta(P-E)
(\frac{1}{(E-P)^2}-\frac{1}{(E+P)^2})
\frac{2T}{P}\log\left(\frac{1-e^{-\frac{E+P}{2T}}}{1-e^{-\frac{P-E}{2T}}}\right)\hspace{0.4cm}\Big] ,
\end{split}\end{equation}
\begin{equation}\begin{split}
F_{-}
=&\frac{\lambda^2}{32\pi P}f(E,T)\times\\
&\hspace{0.4cm}\Big[\ \theta(E-P)
(\frac{1}{(E-P)^2}-\frac{1}{(E+P)^2})
(1+\frac{2T}{P}\log\left(\frac{1-e^{-\frac{E+P}{2T}}}{1-e^{-\frac{E-P}{2T}}}\right))\\
&\hspace{0.8cm}+\theta(P-E)
(\frac{1}{(E-P)^2}-\frac{1}{(E+P)^2})
\frac{2T}{P}\log\left(\frac{1-e^{-\frac{E+P}{2T}}}{1-e^{-\frac{P-E}{2T}}}\right)\hspace{0.4cm}\Big] .
\end{split}\end{equation}

We find that the explicit $\tau_c$ dependence disappears in these expressions.
If we focus on the physics at the fixed physical energy scale $E$, it remains the same with
cosmic evolution. It is a very sensible conclusion as we do not expect physics such as particle mass
to change with cosmic evolution. For a fixed $\varepsilon$, the physical energy $E$ decreases with time
evolution. So the cosmic evolution is identical to the evolution under the renormalization group.
We recall here that the radial coordinate in AdS space corresponds to the energy scale in AdS/CFT
correspondence. Since the radial coordinate in AdS space is related to the time coordinate in de Sitter space
by analytic continuation, de Sitter space seems to be related to AdS space in this respect.
The only physical time dependence appears through the temperature $T$ as it cools down linearly with $\tau_c$
for a fixed $\beta$. 

We do find a non-trivial modification of the distribution function from the Bose distribution due to a large
IR effect.
The effect of the interaction on the distribution function $(\ref{f})$ is such that it 
reduces the particle density in comparison to the Bose distribution.
This effect can be understood as follows. A single particle can turn into two particles due to the cubic interaction.
So such off-shell two particle states are created while the on-shell state weight is reduced by the same amount
due to unitarity. The off-shell states cost more energy and so are less numerous due to the Bose distribution 
function. The net effect is the further reduction of the particle density.

In this section, we have investigated the effects of the interaction on the propagator well
inside the cosmological horizon. The spectral weight of the off-shell states increases with
time while the weight of the on-shell states decreases due to the interaction. 
The modification of the 
Bose distribution is analogous to QCD where the logarithmic divergence requires the scale dependent modification of 
the parton distribution function. 
In term of the physical energy and
momentum variables, explicit time dependence disappears and the time evolution may be identified with
the renormalization group evolution.
So we find that the effects of the interaction in de Sitter space parallel to those in flat space.
As it is explained in Appendix B, these features are also shared by $g\varphi^4$ theory. 
We thus expect they are the universal features of the interacting field theories in de Sitter space.

Nevertheless we should keep in mind that we have investigated the propagator near flat space and the expansion
in terms of $1/p\tau_c$ breaks down near the cosmological horizon.
To fully understand the behavior of the two point function in de Sitter space,
we have to extend our work to the region $|p\tau_c|\sim 1$ and $|p\tau_c|\ll 1$. 

Before concluding this section, we briefly investigate the non-thermal distribution case. 
The modification of the distribution function $\delta f(p,\tau_c)$ by the cubic interaction is roughly of
the following magnitude
\begin{equation}\begin{split}
\frac{\partial f(p,\tau_1,\tau_2)}{\partial \tau_c}
&\sim \frac{\lambda^2}{p}\frac{1}{H^2\tau_c^3}\int_{p+|1/\tau_c|}d\varepsilon\frac{1}{(\varepsilon-p)^2}\\
&\sim \frac{\lambda^2}{p}\frac{1}{H^2\tau_c^2} ,\\
\delta f(p,\tau_1,\tau_2)
&\sim \frac{\lambda^2}{p}\frac{1}{H^2\tau_c} .
\end{split}\end{equation}
So it is $O(1/p|\tau_c|)$ in a generic case instead of $O(1/(p\tau_c)^2)$ for the thermal case.
While it is much larger than the change of the thermal distribution when $p|\tau_c| \gg 1$, 
it becomes only important near the cosmological horizon. 
Although the thermalization  
may take place when the coupling is strong enough $\lambda > H$, it could only occur
near the cosmological horizon. 
At the higher loop level, the thermalization could also take place through the effective $n$ point couplings.
We find this is a very interesting problem which requires further investigations.

\section{Effective cosmological constant}
\setcounter{equation}{0}

\hspace{0.7cm}
In this section, we investigate the energy-momentum tensor of a scalar field.
Since it appears on the right-hand side of the Einstein equation, it provides us an important clue 
to understand the back-reaction to de Sitter space. 
The cosmological constant is renormalized order by order in perturbation theory
as well as other coupling constants such as the Newton's constant. 
If these renormalization effects are time independent, we can assume that they are canceled by
the counter terms. We are thus most interested in whether there is a time dependent back-reaction to the cosmological
constant. 

The energy-momentum tensor for a minimally coupled scalar field is
\begin{equation}\begin{split}
T_{\mu\nu}&\equiv\frac{-2}{\sqrt{-g}}\frac{\delta S_{matter}}{\delta g^{\mu\nu}}\\
&=\ \partial_\mu\varphi\partial_\nu\varphi
-\frac{1}{2}g_{\mu\nu}g^{\alpha\beta}\partial_\alpha\varphi\partial_\beta\varphi
-g_{\mu\nu}\frac{1}{2}m^2\varphi^2
-g_{\mu\nu}\frac{\lambda}{3!}\varphi^3 .
\end{split}\end{equation}
The Einstein equation is
\begin{equation}
R_{\mu\nu}-\frac{1}{2}g_{\mu\nu}R+\Lambda g_{\mu\nu}=\kappa T_{\mu\nu},\hspace{0.8cm}\kappa=8\pi G ,
\label{Einstein}\end{equation}
where $\Lambda$ is the cosmological constant and $G$ is the Newton's constant. 
The effective cosmological constant $\Lambda_{eff}$ may be estimated as
\begin{equation}\begin{split}
\Lambda_{eff}\equiv\ \Lambda-\frac{\kappa}{4}T^{\ \mu}_\mu .
\end{split}\label{effcos}\end{equation}
The vev of $T^{\ \mu}_\mu$ may be evaluated as follows
\begin{equation}\begin{split}
\langle T^{\ \mu}_\mu\rangle 
=&\ \langle -g^{\mu\nu}\partial_\mu\varphi\partial_\nu\varphi
-2m^2\varphi^2-\frac{4\lambda}{3!}\varphi^3\rangle\\
=&-\frac{1}{2}\bigtriangledown^2\langle \varphi^2\rangle
-m^2\langle \varphi^2\rangle
-\frac{\lambda}{6}\langle \varphi^3\rangle ,
\end{split}\label{trace}\end{equation}
where we have used the equation of motion.

Let us investigate the expectation value of the energy-momentum tensor for a free massless field case
first.
We may formally evaluate the energy-momentum tensor as
\begin{eqnarray}
T_{00}&=& \int {d^3p \over (2\pi )^32p}(p^2H^2\tau^2+{1\over 2}H^2)(1+2f(p)) ,\nonumber \\
T_{ii}&=& \int {d^3p \over (2\pi )^32p}({1\over 3}p^2H^2\tau^2-{1\over 6}H^2)(1+2f(p)),~ i=1,2,3 .
\end{eqnarray}
Although they are quartically divergent,
the UV contribution to the renormalized energy-momentum tensor
is determined from the conformal anomaly $T^{\ \mu}_\mu$ 
\cite{DC}\cite{BD} 
\begin{equation}
T_{\mu\nu}={1\over 64\pi^2}{29\over 15}H^4 g_{\mu\nu} .
\end{equation}
We may assume that its contribution to the effective cosmological constant through (\ref{effcos}) is canceled by 
a counter term in the bare cosmological constant.
From the Bose distribution function $f$, we find the following contribution
\begin{eqnarray}
T_{00}&=& ({\pi^2\over 30}T^4+{1\over 24}H^2T^2){1\over (H\tau)^2}={\cal E}{1\over (H\tau)^2} ,\nonumber \\
T_{ii}&=& ({\pi^2\over 90}T^4-{1\over 72}H^2T^2){1\over (H\tau)^2}={\cal P}{1\over (H\tau)^2},~ i=1,2,3 .
\end{eqnarray}
Since the physical temperature $T\sim |\tau|$, the energy ${\cal E}$ and pressure ${\cal P}$ decreases with time and 
the cosmological constant becomes more and more dominant once $T<H$.

\subsection{Contribution from inside the cosmological horizon}

\hspace{0.7cm}
Firstly, we evaluate the contribution to the energy-momentum tensor from the degrees of freedom
well inside the cosmological horizon $|p\tau|\gg 1$. 
From the two point function, we can estimate the trace of the energy-momentum tensor as in (\ref{trace}).
We substitute the full propagator specified by (\ref{f}), (\ref{Z}), (\ref{F+}), (\ref{F-}) 
into $-\frac{1}{2}\bigtriangledown^2\langle \varphi^2\rangle$
\begin{equation}\begin{split}
-\frac{1}{2}\bigtriangledown^2\langle \varphi^2\rangle
=\int\frac{d^3p}{(2\pi)^32p}\frac{3\lambda^2H^2}{64\pi^2p}\frac{\beta}{e^{\beta p}-1}\frac{e^{\beta p}}{e^{\beta p}-1} .
\end{split}\end{equation}
Of course, we can obtain the same result without solving the Boltzmann equation. 
We have checked that the following expression in Schwinger-Keldysh perturbation theory
reproduces the identical result
\begin{equation}\begin{split}
\langle \varphi^2(x)\rangle
=\langle in|\left(\frac{-i\lambda}{3!}\right)^2
\int^\tau_{-\infty}\frac{d\tau_1}{H^4\tau_1^4}
\int^{\tau_1}_{-\infty}\frac{d\tau_2}{H^4\tau_2^4}\int d^3x_1d^3x_2\ 
\big[[\varphi^2(x),\varphi^3(x_1)],\varphi^3(x_2)\big]\ |in\rangle .
\end{split}\end{equation}

We also need to estimate the contribution from the three point function in (\ref{trace}). 
It turns out that the three point function does not contribute to the effective cosmological constant
as it is $O(\lambda^2T^2)$. See Appendix C for the details.
So the contribution to the effective cosmological constant from the degrees of freedom inside the cosmological
horizon is  
\begin{equation}\begin{split}
-{\kappa\over 4}\langle T_\mu^{\ \mu}\rangle
&=-{\kappa\over 4}\int_{|1/\tau|}\frac{d^3p}{(2\pi)^32p}\frac{3\lambda^2H^2}{64\pi^2p}\frac{\beta}{e^{\beta p}-1}
\frac{e^{\beta p}}{e^{\beta p}-1}\\
&=\ -{\kappa}\frac{3\lambda^2H^2}{2^{10}\pi^4}\frac{1}{e^{H/T}-1} .
\end{split}\end{equation}
Although we find that it reduces the original cosmological constant, 
it vanishes as the universe cools down with time.
It therefore does not lead to a decreasing cosmological constant.
In order to find such an effect,
we may need to consider the contributions to the cosmological constant from the degrees of freedom
outside the cosmological horizon as follows.

\subsection{Contribution from outside the cosmological horizon}

\hspace{0.7cm}
In this subsection, we calculate the contribution to the energy-momentum tensor from 
the degrees of freedom 
outside the cosmological horizon: $|p\tau|\ll 1$.
Inside the cosmological horizon, the degrees of freedom
are constant because we adopt a fixed physical UV cut-off $\Lambda$ and IR cut-off $H$
corresponding to the following energy integration range
\begin{equation}\begin{split}
\int^{\Lambda/H|\tau|}_{1/|\tau|}\frac{d\varepsilon}{2\pi} .
\end{split}\end{equation}
On the other hand 
the degrees of freedom outside the cosmological horizon increase as time goes on.
They correspond to the following energy integration range
\begin{equation}\begin{split}
\int^{1/|\tau|}_{\varepsilon_0}\frac{d\varepsilon}{2\pi} ,
\end{split}\end{equation}
where $\varepsilon_0$ denotes an IR cut-off. 
Simply put, the degrees of freedom outside the cosmological horizon increase as more and more degrees of
freedom go out of the cosmological horizon with cosmic evolution in de Sitter space. 
Since every degree of freedom can contribute to the cosmological constant, they may give rise to an accumulating effect.
In fact we find that $\langle T_\mu^{\ \mu}\rangle$ receives a time dependent contribution from the degrees of freedom 
outside the cosmological horizon. 

To investigate this effect through the trace of the energy-momentum tensor (\ref{trace}), 
we need to calculate the expectation value of the two point and the 
three point functions. 
As it is explained in Appendix C, $\langle T_\mu^{\ \mu}\rangle$ is estimated 
for the $f=0$ case  as follows
\begin{equation}\begin{split}
\langle T_\mu^{\ \mu}\rangle
=-\frac{1}{2}\bigtriangledown^2\langle \varphi^2\rangle-\frac{\lambda}{6}\langle\varphi^3\rangle
=\ \frac{1}{2^3\cdot 3^2}\ \frac{\lambda^2H^2}{\pi^4}(-\log (-\varepsilon_0\tau))^3>0 .
\end{split}\end{equation}
We thus find that the contribution from the degrees of freedom outside the cosmological horizon 
gives rise to a very interesting effect.
It screens the cosmological constant as the universe evolves 
\begin{equation}\begin{split}
\Lambda_{eff}=\Lambda-
\frac{\kappa\lambda^2H^2}{2^5\cdot 3^2\pi^4}(-\log (-\varepsilon_0\tau))^3 .
\end{split}\end{equation}

In this paper we have investigated quantum effects which diminish the cosmological constant.
Of course a scalar field can classically roll down a potential to lower the effective cosmological constant.
Therefore it is interesting to ask whether quantum screening effects exist in classically
stable theory such as  $g\varphi^4$ theory.
In this case, $\langle T_\mu^{\ \mu}\rangle$ is given by a two point function 
after using the equation of motion 
\begin{align}
\langle T^{\ \mu}_\mu\rangle 
=-\frac{1}{2}\bigtriangledown^2\langle \varphi^2\rangle.
\end{align}
The expectation value of the two point function is 
\begin{align}
\langle \varphi^2\rangle
=\frac{g^2H^2}{2^6\cdot 3^3\cdot 5\pi^6}(-\log(-\varepsilon_0\tau))^5.
\end{align}
The effective cosmological constant is estimated as
\begin{align}
\Lambda_{eff}=\Lambda-\frac{\kappa g^2H^4}{2^9\cdot3^2\pi^6}(-\log(-\varepsilon_0\tau))^4.
\end{align}
It turns out that the cosmological constant is also screened in $g\varphi^4$ theory
in an analogous way .  Our findings in this subsection are in agreement with  \cite{TWM}.

\section{Conclusions}
\setcounter{equation}{0}

\hspace{0.7cm}
In this paper, we have investigated an interacting scalar field theory in
de Sitter space. As Feynman-Dyson perturbation theory breaks down, we need to employ Schwinger-Keldysh
formalism. Our problem therefore belong to non-equilibrium physics. 
It may illuminates great mysteries of our time: inflation in the early universe and dark energy today.
Phenomenologically it appears that cosmological constant is not a constant but it changes with cosmic
evolution. 
Since the vacuum changes with time in de Sitter space, it is logically possible that
the effective cosmological constant also changes with time. It is because 
the expectation value of the matter energy-momentum tensor
contributes to the effective cosmological constant through the Einstein's equation of motion.
We are thus most interested in a possible time dependence of the effective cosmological constant.

We have first investigated the time dependence of the propagator well inside the cosmological horizon. 
We have derived a Boltzmann equation from
a Schwinger-Dyson equation. We have found that the total integral of the spectral weight remains to be unity
as the particle creation effects are accompanied by the reduction of the on-shell states.
The leading IR contributions cancel between the real and virtual processes.
This fact suffices for the complete cancellation of the infra-red divergences at zero temperature as pointed out in section 4.
However it is not so at finite temperature. 
The remaining IR contribution leads to the modification of the particle distribution function as (\ref{mdf}).
This interesting effect vanishes in the zero temperature limit as the Bose distribution function itself vanishes.

Although the propagator is time dependent, explicit time dependence disappears when it is expressed
by physical quantities. In another words, time evolution may be identified with the renormalization
group evolution in an interacting field theory in de Sitter space.
We have thus found that a field theory in de Sitter space shares many common properties with
those in Minkowski space. We believe this is due to unitarity of the theory. We also believe
it is due to the fact that the degrees of freedom inside the cosmological horizon does
not change under cosmic evolution. In any field theory we need to adopt a fixed physical UV cut-off
$\Lambda$ while IR cut-off is provided by the Hubble constant.
From this reason, we find that the degrees of freedom inside the cosmological horizon do not
lead to diminishing cosmological constant.

On the other hand, the degrees of freedom outside the cosmological horizon increase with
time as more and more degrees go out of the horizon. Indeed we find that they may give rise to a 
desired effect. 
As they accumulate with cosmic evolution, the screening effect of the cosmological constant 
grows. Although it is suppressed by $\kappa\lambda^2$, the suppression effect
can be compensated by large logarithmic factors as arbitrary many degrees of freedom could
accumulate outside the horizon.\footnote{Such a mechanism is proposed by Tsamis and Woodard. See \cite{TWM2} and
references therein.}
It is important to generalize our work to more generic situations including quantum effects of gravity.
Since our investigation of the propagators is limited to the region well inside the horizon,
we also need to understand the behavior of them around and beyond the cosmological horizon.

\section*{Acknowledgments}

\hspace{0.7cm}
This work is supported in part by Grant-in-Aid for Scientific Research from
the Ministry of Education, Science and Culture of Japan.
We would like to thank A. Ishibashi, S. Iso, K. Itakura, H. Kodama, E. Komatsu and H. Suzuki for discussions
and information.

\appendix

\section{Collision term evaluation}
\setcounter{equation}{0}

\hspace{0.7cm}
In this appendix we explain the details of our calculation for the collision term.

In the first step, using our integration formula (\ref{Gamma}), 
the on-shell collision term (\ref{origin1}) is evaluated as
\begin{align}
&C_{on}[f]\notag\\
=&+(1+f(p))e^{-ip\bar{\tau}}\times\frac{1}{H^2}\frac{(-i\lambda)^2}{2}\frac{1}{32\pi^2p^2}
\int^\infty_0dp_1\int^{p_1+p}_{|p_1-p|}dp_2\\
&\hspace{0.8cm}\times\Big[
+\big\{\frac{1}{i(p_1+p_2-p)}\frac{-2\bar{\tau}}{\tau_c^3}
+\frac{-1}{(p_1+p_2-p)^2}\frac{2}{\tau_c^3}\big\}\notag\\
&\hspace{4cm}\times\big\{(1+f(p_1))(1+f(p_2))-f(p_1)f(p_2)\big\}\notag\\
&\hspace{1.5cm}+\big\{\frac{1}{i(p_1-p_2-p)}\frac{-2\bar{\tau}}{\tau_c^3}
+\frac{-1}{(p_1-p_2-p)^2}\frac{2}{\tau_c^3}\big\}\notag\\
&\hspace{4cm}\times\big\{(1+f(p_1))f(p_2)-f(p_1)(1+f(p_2))\big\}\notag\\
&\hspace{1.5cm}+\big\{\frac{1}{i(-p_1+p_2-p)}\frac{-2\bar{\tau}}{\tau_c^3}
+\frac{-1}{(-p_1+p_2-p)^2}\frac{2}{\tau_c^3}\big\}\notag\\
&\hspace{4cm}\times\big\{f(p_1)(1+f(p_2))-(1+f(p_1))f(p_2)\big\}\notag\\
&\hspace{1.5cm}+\big\{\frac{1}{i(-p_1-p_2-p)}\frac{-2\bar{\tau}}{\tau_c^3}
+\frac{-1}{(-p_1-p_2-p)^2}\frac{2}{\tau_c^3}\big\}\notag\\
&\hspace{4cm}\times\big\{f(p_1)f(p_2)-(1+f(p_1))(1+f(p_2))\big\}\ \Big]\notag\\
&-f(p)\ e^{+ip\bar{\tau}}\times\frac{1}{H^2}\frac{(-i\lambda)^2}{2}\frac{1}{32\pi^2p^2}
\int^\infty_0dp_1\int^{p_1+p}_{|p_1-p|}dp_2\notag\\
&\hspace{0.8cm}\times\Big[
+\big\{\frac{1}{i(p_1+p_2-p)}\frac{+2\bar{\tau}}{\tau_c^3}
+\frac{-1}{(p_1+p_2-p)^2}\frac{2}{\tau_c^3}\big\}\notag\\
&\hspace{4cm}\times\big\{(1+f(p_1))(1+f(p_2))-f(p_1)f(p_2)\big\}\notag\\
&\hspace{1.5cm}+\big\{\frac{1}{i(p_1-p_2-p)}\frac{+2\bar{\tau}}{\tau_c^3}
+\frac{-1}{(p_1-p_2-p)^2}\frac{2}{\tau_c^3}\big\}\notag\\
&\hspace{4cm}\times\big\{(1+f(p_1))f(p_2)-f(p_1)(1+f(p_2))\big\}\notag\\
&\hspace{1.5cm}+\big\{\frac{1}{i(-p_1+p_2-p)}\frac{+2\bar{\tau}}{\tau_c^3}
+\frac{-1}{(-p_1+p_2-p)^2}\frac{2}{\tau_c^3}\big\}\notag\\
&\hspace{4cm}\times\big\{f(p_1)(1+f(p_2))-(1+f(p_1))f(p_2)\big\}\notag\\
&\hspace{1.5cm}+\big\{\frac{1}{i(-p_1-p_2-p)}\frac{+2\bar{\tau}}{\tau_c^3}
+\frac{-1}{(-p_1-p_2-p)^2}\frac{2}{\tau_c^3}\big\}\notag\\
&\hspace{4cm}\times\big\{f(p_1)f(p_2)-(1+f(p_1))(1+f(p_2))\big\}\ \Big].\notag
\end{align}
Here we have used the following relation.
\begin{equation}\begin{split}
&\frac{1}{2p}\int\frac{d^3p_1}{(2\pi)^32p_1}\frac{d^3p_2}{(2\pi)^32p_2}(2\pi)^3\delta^{(3)}
({\bf p}_1+{\bf p}_2-{\bf p})\\
=&\frac{1}{32\pi^2p^2}\int^\infty_0dp_1\int^{p_1+p}_{|p_1-p|}dp_2 .
\end{split}\end{equation}
For the comparison with the off-sell part, we insert the identity factor as
\begin{equation}
\int \frac{d\varepsilon}{2\pi}\ (2\pi)\delta(\varepsilon-(\pm p_1\pm p_2)) .
\label{delta}\end{equation}
In this way, we obtain the expression (\ref{on}) in the main text.
The off-shell part is calculated just like the on-shell part.  

In the main text, we have introduced the collision terms with a finite energy resolution $\Delta\varepsilon$
following a standard procedure in massless field theories.
They are given explicitly as follows
\begin{align}
C_{on}'[f]
\equiv &\ C_{on}[f]\notag\\
&+\frac{\lambda^2}{16\pi p^2H^2}\times\label{on_1}\\
&\hspace{0.4cm}\Big[\int^{p+\Delta\varepsilon}_p\frac{d\varepsilon}{2\pi}\ e^{-i\varepsilon\bar{\tau}}
(\frac{1}{(\varepsilon-p)^2}-\frac{1}{(\varepsilon+p)^2})\frac{-1}{\tau_c^3}
\int^{\frac{\varepsilon+p}{2}}_{\frac{\varepsilon-p}{2}}dp_1\ (1+f(p_1))(1+f(\varepsilon-p_1))\notag\\
&\hspace{0.4cm}+2\int^p_{p-\Delta\varepsilon}\frac{d\varepsilon}{2\pi}\ e^{-i\varepsilon\bar{\tau}}
(\frac{1}{(\varepsilon-p)^2}-\frac{1}{(\varepsilon+p)^2})\frac{-1}{\tau_c^3}
\int^\infty_{\frac{\varepsilon+p}{2}}dp_1\ (1+f(p_1))f(p_1-\varepsilon)\hspace{0.8cm}\Big]\notag\\
&-\frac{\lambda^2}{16\pi p^2H^2}\times\notag\\
&\hspace{0.4cm}\Big[\int^{p+\Delta\varepsilon}_p\frac{d\varepsilon}{2\pi}\ e^{+i\varepsilon\bar{\tau}}
(\frac{1}{(\varepsilon-p)^2}-\frac{1}{(\varepsilon+p)^2})\frac{-1}{\tau_c^3}
\int^{\frac{\varepsilon+p}{2}}_{\frac{\varepsilon-p}{2}}dp_1\ f(p_1)f(\varepsilon-p_1)\notag\\
&\hspace{0.4cm}+2\int^p_{p-\Delta\varepsilon}\frac{d\varepsilon}{2\pi}\ e^{+i\varepsilon\bar{\tau}}
(\frac{1}{(\varepsilon-p)^2}-\frac{1}{(\varepsilon+p)^2})\frac{-1}{\tau_c^3}
\int^\infty_{\frac{\varepsilon+p}{2}}dp_1\ f(p_1)(1+f(p_1-\varepsilon))\ \Big],\notag
\end{align}
\begin{align}
&C_{off}'[f]\notag\\
=&+\frac{\lambda^2}{16\pi p^2H^2}\times\label{off_1}\\
&\hspace{0.4cm}\Big[\int^\infty_{p+\Delta\varepsilon}\frac{d\varepsilon}{2\pi}\ e^{-i\varepsilon\bar{\tau}}
(\frac{1}{(\varepsilon-p)^2}-\frac{1}{(\varepsilon+p)^2})\frac{-1}{\tau_c^3}
\int^{\frac{\varepsilon+p}{2}}_{\frac{\varepsilon-p}{2}}dp_1\ (1+f(p_1))(1+f(\varepsilon-p_1))\notag\\
&\hspace{0.4cm}+2\int^{p-\Delta\varepsilon}_0\frac{d\varepsilon}{2\pi}\ e^{-i\varepsilon\bar{\tau}}
(\frac{1}{(\varepsilon-p)^2}-\frac{1}{(\varepsilon+p)^2})\frac{-1}{\tau_c^3}
\int^\infty_{\frac{\varepsilon+p}{2}}dp_1\ (1+f(p_1))f(p_1-\varepsilon)\hspace{0.4cm}\Big]\notag\\
&-\frac{\lambda^2}{16\pi p^2H^2}\times\notag\\
&\hspace{0.4cm}\Big[\int^\infty_{p+\Delta\varepsilon}\frac{d\varepsilon}{2\pi}\ e^{+i\varepsilon\bar{\tau}}
(\frac{1}{(\varepsilon-p)^2}-\frac{1}{(\varepsilon+p)^2})\frac{-1}{\tau_c^3}
\int^{\frac{\varepsilon+p}{2}}_{\frac{\varepsilon-p}{2}}dp_1\ f(p_1)f(\varepsilon-p_1)\notag\\
&\hspace{0.4cm}+2\int^{p-\Delta\varepsilon}_0\frac{d\varepsilon}{2\pi}\ e^{+i\varepsilon\bar{\tau}}
(\frac{1}{(\varepsilon-p)^2}-\frac{1}{(\varepsilon+p)^2})\frac{-1}{\tau_c^3}
\int^\infty_{\frac{\varepsilon+p}{2}}dp_1\ f(p_1)(1+f(p_1-\varepsilon))\ \Big].\notag
\end{align}

In the case of the thermal distribution function, 
the on-shell collision term (\ref{on_1}) is evaluated as  
\begin{equation*}\begin{split}
&C_{on}'[f]\\
=&-\frac{\lambda^2}{16\pi pH^2}(1+f(p))e^{-ip\bar{\tau}}\times\\
&\hspace{0.4cm}\Big[\big(\int^\infty_{p+\Delta\varepsilon}+\int^{p+\Delta\varepsilon}_p\big)
\frac{d\varepsilon}{2\pi}
\big\{(\frac{1}{\varepsilon-p}+\frac{1}{\varepsilon+p})\frac{i\bar{\tau}}{\tau_c^3}
+(\frac{1}{(\varepsilon-p)^2}-\frac{1}{(\varepsilon+p)^2})\frac{-1}{\tau_c^3}\big\}(1+ G(\varepsilon, p, \beta )
)\\
&\hspace{0.8cm}+\big(\int^p_{p-\Delta\varepsilon}+\int^{p-\Delta\varepsilon}_0\big)
\frac{d\varepsilon}{2\pi}
\big\{(\frac{1}{\varepsilon-p}+\frac{1}{\varepsilon+p})\frac{i\bar{\tau}}{\tau_c^3}
+(\frac{1}{(\varepsilon-p)^2}-\frac{1}{(\varepsilon+p)^2})\frac{-1}{\tau_c^3}\big\}G(\varepsilon, p, \beta )
\Big]\\
&+\frac{\lambda^2}{16\pi pH^2}\ f(p)e^{+ip\bar{\tau}}\times\\
&\hspace{0.4cm}\Big[\big(\int^\infty_{p+\Delta\varepsilon}+\int^{p+\Delta\varepsilon}_p\big)
\frac{d\varepsilon}{2\pi}
\big\{(\frac{1}{\varepsilon-p}+\frac{1}{\varepsilon+p})\frac{-i\bar{\tau}}{\tau_c^3}
+(\frac{1}{(\varepsilon-p)^2}-\frac{1}{(\varepsilon+p)^2})\frac{-1}{\tau_c^3}\big\}
(1+G(\varepsilon, p, \beta )
)\\
&\hspace{0.8cm}+\big(\int^p_{p-\Delta\varepsilon}+\int^{p-\Delta\varepsilon}_0\big)
\frac{d\varepsilon}{2\pi}
\big\{(\frac{1}{\varepsilon-p}+\frac{1}{\varepsilon+p})\frac{-i\bar{\tau}}{\tau_c^3}
+(\frac{1}{(\varepsilon-p)^2}-\frac{1}{(\varepsilon+p)^2})\frac{-1}{\tau_c^3}\big\}
G(\varepsilon, p, \beta )
\Big]
\end{split}\end{equation*}
\begin{equation}\begin{split}
\ \ &+\frac{\lambda^2}{16\pi pH^2}\times\\
&\hspace{0.4cm}\Big[\int^{p+\Delta\varepsilon}_p\frac{d\varepsilon}{2\pi}\ (1+f(\varepsilon))e^{-i\varepsilon\bar{\tau}}
(\frac{1}{(\varepsilon-p)^2}-\frac{1}{(\varepsilon+p)^2})\frac{-1}{\tau_c^3}
(1+G(\varepsilon, p, \beta )
)\\
&\hspace{0.8cm}+\int^p_{p-\Delta\varepsilon}\frac{d\varepsilon}{2\pi}\ (1+f(\varepsilon))e^{-i\varepsilon\bar{\tau}}
(\frac{1}{(\varepsilon-p)^2}-\frac{1}{(\varepsilon+p)^2})\frac{-1}{\tau_c^3}
G(\varepsilon, p, \beta )
\hspace{0.4cm}\Big]\\
&-\frac{\lambda^2}{16\pi pH^2}\times\\
&\hspace{0.4cm}\Big[\int^{p+\Delta\varepsilon}_p\frac{d\varepsilon}{2\pi}\ f(\varepsilon)e^{+i\varepsilon\bar{\tau}}
(\frac{1}{(\varepsilon-p)^2}-\frac{1}{(\varepsilon+p)^2})\frac{-1}{\tau_c^3}
(1+G(\varepsilon, p, \beta )
)\\
&\hspace{0.8cm}+\int^p_{p-\Delta\varepsilon}\frac{d\varepsilon}{2\pi}\ f(\varepsilon)e^{+i\varepsilon\bar{\tau}}
(\frac{1}{(\varepsilon-p)^2}-\frac{1}{(\varepsilon+p)^2})\frac{-1}{\tau_c^3}
G(\varepsilon, p, \beta ) 
\hspace{0.4cm}\Big] .
\end{split}\label{on_2}\end{equation}
$G(\varepsilon, p, \beta )$ is defined in (\ref{Gfn}).
We find that the linear infra-red divergences at $\varepsilon=p$ are canceled, but the 
apparent logarithmic divergences remain. 
The situation here is analogous to QCD where the logarithmic divergences require the scale dependent modification of 
the parton distribution function. In our case, the IR singularity also leads to the modification of the particle
distribution function as the final expression is shown in the main text (\ref{on_3}).

\section{Boltzmann equation in $g\varphi^4$ theory}
\setcounter{equation}{0}

\hspace{0.7cm}
In this appendix, we consider the Boltzmann equation in $g\varphi^4$ theory. 
Since this theory is classically stable,  
it is a good example for investigating quantum effects on the dS background.

As in the main text, we evaluate the time integrations with the assumption $|(\varepsilon\pm p)\tau_i|\gg 1$. 
In $g\varphi^4$ theory, we need to retain higher order terms than (\ref{Gamma}) to
investigate the particle production effects in de Sitter space 
\begin{equation}\begin{split}
\int^{\tau_i}_{-\infty}d\tau_3\ \frac{1}{\tau_3^n}e^{i(\varepsilon\pm p)\tau_3}
\sim \ e^{i(\varepsilon\pm p)\tau_i}\times
\left[\frac{1}{i(\varepsilon\pm p)\tau_i^n}+\frac{-n}{(\varepsilon\pm p)^2\tau_i^{n+1}}
+\frac{-n(n+1)}{i(\varepsilon\pm p)^3\tau_i^{n+2}}\right] ,&\\
n=1,2,\cdots .&
\label{Gamma2}\end{split}\end{equation}
We should note that (\ref{Gamma2}) can be evaluated exactly when $n=0$
\begin{eqnarray}
\int^{\tau_i}_{-\infty}d\tau_3\ e^{i(\varepsilon\pm p)\tau_3}
&=& \ e^{i(\varepsilon\pm p)\tau_i}\times
\frac{1}{i(\varepsilon\pm p-i0)} \\ \nonumber
&=& \ e^{i(\varepsilon\pm p)\tau_i}\times
\left({P \over i(\varepsilon\pm p)}+\pi\delta (\varepsilon\pm p)\right) .
\label{n=0}
\end{eqnarray}
The $-i0$ prescription is necessary for the convergence at $\tau_3=-\infty$. 

In this appendix, we focus on the IR effects of the collision term at $\varepsilon - p= 0$. 
Therefore we consider only $2 \rightarrow 2$ processes. 
In these processes, the on-shell part of the collision term is as follows
\begin{align}
C_{on}[f]=
&+(1+f(p))\ e^{-ip\bar{\tau}}\times\frac{(-ig)^2}{6}\frac{1}{2p}\int\prod^3_{i=1}\frac{d^3p_i}{(2\pi)^32p_i}
\ (2\pi)^3\delta^{(3)}({\bf p}+{\bf p}_1+{\bf p}_2+{\bf p}_3)\notag\\
&\hspace{0.4cm}\times \Big[+3\big\{(1+f(p_1))(1+f(p_2))f(p_3)-f(p_1)f(p_2)(1+f(p_3))\big\}\notag\\
&\hspace{2.4cm}\times\big\{+2\pi\delta(p_1+p_2-p_3-p)\label{on4}\\
&\hspace{3.2cm}+\frac{1}{i(p_1+p_2-p_3-p)}\times(\frac{1}{p_1^2}+\frac{1}{p_2^2}+\frac{1}{p_3^2})\frac{-2\bar{\tau}}{\tau_c^3}\notag\\
&\hspace{3.2cm}+\frac{1}{i(p_1+p_2-p_3-p)^2}\times(\frac{1}{p_1}+\frac{1}{p_2}-\frac{1}{p_3}-\frac{1}{p})\frac{-2\bar{\tau}}{\tau_c^3}\notag\\
&\hspace{3.2cm}+\frac{1}{(p_1+p_2-p_3-p)^2}\times(\frac{1}{p_1^2}+\frac{1}{p_2^2}+\frac{1}{p_3^2}+\frac{1}{p^2})\frac{-2}{\tau_c^3}\notag\\
&\hspace{3.2cm}+\frac{1}{(p_1+p_2-p_3-p)^3}\times(\frac{1}{p_1}+\frac{1}{p_2}-\frac{1}{p_3}-\frac{1}{p})\frac{-4}{\tau_c^3}\hspace{0.4cm}\big\}\ \Big]\notag\\
&-f(p)\ e^{+ip\bar{\tau}}\times\frac{(-ig)^2}{6}\frac{1}{2p}\int\prod^3_{i=1}\frac{d^3p_i}{(2\pi)^32p_i}
\ (2\pi)^3\delta^{(3)}({\bf p}+{\bf p}_1+{\bf p}_2+{\bf p}_3)\notag\\
&\hspace{0.4cm}\times \Big[+3\big\{(1+f(p_1))(1+f(p_2))f(p_3)-f(p_1)f(p_2)(1+f(p_3))\big\}\notag\\
&\hspace{2.4cm}\times\big\{+2\pi\delta(p_1+p_2-p_3-p)\notag\\
&\hspace{3.2cm}-\frac{1}{i(p_1+p_2-p_3-p)}\times(\frac{1}{p_1^2}+\frac{1}{p_2^2}+\frac{1}{p_3^2})\frac{-2\bar{\tau}}{\tau_c^3}\notag\\
&\hspace{3.2cm}-\frac{1}{i(p_1+p_2-p_3-p)^2}\times(\frac{1}{p_1}+\frac{1}{p_2}-\frac{1}{p_3}-\frac{1}{p})\frac{-2\bar{\tau}}{\tau_c^3}\notag\\
&\hspace{3.2cm}+\frac{1}{(p_1+p_2-p_3-p)^2}\times(\frac{1}{p_1^2}+\frac{1}{p_2^2}+\frac{1}{p_3^2}+\frac{1}{p^2})\frac{-2}{\tau_c^3}\notag\\
&\hspace{3.2cm}+\frac{1}{(p_1+p_2-p_3-p)^3}\times(\frac{1}{p_1}+\frac{1}{p_2}-\frac{1}{p_3}-\frac{1}{p})\frac{-4}{\tau_c^3}\hspace{0.4cm}\big\}\ \Big]. \notag
\end{align}

The off-shell part of collision term is as follows
\begin{align}
C_{off}[f]=
&-\frac{(-ig)^2}{6}\frac{1}{2p}\int\prod^3_{i=1}\frac{d^3p_i}{(2\pi)^32p_i}
\ (2\pi)^3\delta^{(3)}({\bf p}+{\bf p}_1+{\bf p}_2+{\bf p}_3)\notag\\
&\hspace{0.4cm}\times \Big[+3(1+f(p_1))(1+f(p_2))f(p_3)\ e^{-i(p_1+p_2-p_3)\bar{\tau}}\notag\\
&\hspace{2.4cm}\times\big\{+2\pi\delta(p_1+p_2-p_3-p)\label{off4}\\
&\hspace{3.2cm}-\frac{1}{i(p_1+p_2-p_3-p)}\times\frac{1}{p^2}\frac{-2\bar{\tau}}{\tau_c^3}\notag\\
&\hspace{3.2cm}-\frac{1}{i(p_1+p_2-p_3-p)^2}\times(\frac{1}{p_1}+\frac{1}{p_2}-\frac{1}{p_3}-\frac{1}{p})\frac{-2\bar{\tau}}{\tau_c^3}\notag\\
&\hspace{3.2cm}+\frac{1}{(p_1+p_2-p_3-p)^2}\times(\frac{1}{p_1^2}+\frac{1}{p_2^2}+\frac{1}{p_3^2}+\frac{1}{p^2})\frac{-2}{\tau_c^3}\notag\\
&\hspace{3.2cm}+\frac{1}{(p_1+p_2-p_3-p)^3}\times(\frac{1}{p_1}+\frac{1}{p_2}-\frac{1}{p_3}-\frac{1}{p})\frac{-4}{\tau_c^3}\hspace{0.4cm}\big\}\ \Big]\notag\\
&+\frac{(-ig)^2}{6}\frac{1}{2p}\int\prod^3_{i=1}\frac{d^3p_i}{(2\pi)^32p_i}
\ (2\pi)^3\delta^{(3)}({\bf p}+{\bf p}_1+{\bf p}_2+{\bf p}_3)\notag\\
&\hspace{0.4cm}\times \Big[+3f(p_1)f(p_2)(1+f(p_3))\ e^{+i(p_1+p_2-p_3)\bar{\tau}}\notag\\
&\hspace{2.4cm}\times\big\{+2\pi\delta(p_1+p_2-p_3-p)\notag\\
&\hspace{3.2cm}+\frac{1}{i(p_1+p_2-p_3-p)}\times\frac{1}{p^2}\frac{-2\bar{\tau}}{\tau_c^3}\notag\\
&\hspace{3.2cm}+\frac{1}{i(p_1+p_2-p_3-p)^2}\times(\frac{1}{p_1}+\frac{1}{p_2}-\frac{1}{p_3}-\frac{1}{p})\frac{-2\bar{\tau}}{\tau_c^3}\notag\\
&\hspace{3.2cm}+\frac{1}{(p_1+p_2-p_3-p)^2}\times(\frac{1}{p_1^2}+\frac{1}{p_2^2}+\frac{1}{p_3^2}+\frac{1}{p^2})\frac{-2}{\tau_c^3}\notag\\
&\hspace{3.2cm}+\frac{1}{(p_1+p_2-p_3-p)^3}\times(\frac{1}{p_1}+\frac{1}{p_2}-\frac{1}{p_3}-\frac{1}{p})\frac{-4}{\tau_c^3}\hspace{0.4cm}\big\}\ \Big]. \notag
\end{align}
In (\ref{on4}) and (\ref{off4}), only the leading term in $1/p|\tau_c|$  expansion is shown 
for the energy conserving part
containing $\delta(p_1+p_2-p_3-p)$. 

We note that the leading term in $1/p|\tau_c|$  expansion is the same with the collision term in Minkowski space
\begin{align}
&C[f]_{\ leading}\notag\\
=\ &\frac{g^2}{2}\frac{1}{2p}\int\prod^3_{i=1}\frac{d^3p_i}{(2\pi)^32p_i}
\ (2\pi)^4\delta^{(3)}({\bf p}+{\bf p}_1+{\bf p}_2+{\bf p}_3) \delta(p_1+p_2-p_3-p)\\
&\times\Big[+\{f(p_1)f(p_2)(1+f(p_3))(1+f(p))-(1+f(p_1))(1+f(p_2))f(p_3)f(p)\}\ e^{-ip\bar{\tau}}\notag\\
&\hspace{0.8cm}-\{f(p_1)f(p_2)(1+f(p_3))(1+f(p))-(1+f(p_1))(1+f(p_2))f(p_3)f(p)\}\ e^{+ip\bar{\tau}}\ \Big]. \notag
\end{align}
This is because the leading term is conformally invariant. 
We thus obtain the identical result with  \cite{Hohenegger} to the leading order in $1/p|\tau_c|$  expansion.

In addition to the leading effect, we investigate the particle production effects due to energy non-conservation. 
Let us focus on the case that the initial distribution function is thermal. 
It solves the Boltzmann equation to the leading order as the following identity holds
\begin{equation}\begin{split}
&(1+f(p_1))(1+f(p_2))f(p_3)f(p_1+p_2-p_3)\\
=\ &f(p_1)f(p_2)(1+f(p_3))(1+f(p_1+p_2-p_3)).
\end{split}\end{equation}
Therefore the off-shell part is written as follows
\begin{align}
&C_{off}[f]_{\ next\ leading}\notag\\=
&-\frac{(-ig)^2}{6}\frac{1}{2p}\int\prod^3_{i=1}\frac{d^3p_i}{(2\pi)^32p_i}
\ (2\pi)^3\delta^{(3)}({\bf p}+{\bf p}_1+{\bf p}_2+{\bf p}_3)\\
&\hspace{0.4cm}\times \Big[+3\{(1+f(p_1))(1+f(p_2))f(p_3)-f(p_1)f(p_2)(1+f(p_3))\}\notag\\
&\hspace{2.4cm}\times(1+f(p_1+p_2-p_3))\ e^{-i(p_1+p_2-p_3)\bar{\tau}}\notag\\
&\hspace{2.4cm}\times\big\{
-\frac{1}{i(p_1+p_2-p_3-p)}\times\frac{1}{p^2}\frac{-2\bar{\tau}}{\tau_c^3}\notag\\
&\hspace{3.2cm}-\frac{1}{i(p_1+p_2-p_3-p)^2}\times(\frac{1}{p_1}+\frac{1}{p_2}-\frac{1}{p_3}-\frac{1}{p})\frac{-2\bar{\tau}}{\tau_c^3}\notag\\
&\hspace{3.2cm}+\frac{1}{(p_1+p_2-p_3-p)^2}\times(\frac{1}{p_1^2}+\frac{1}{p_2^2}+\frac{1}{p_3^2}+\frac{1}{p^2})\frac{-2}{\tau_c^3}\notag\\
&\hspace{3.2cm}+\frac{1}{(p_1+p_2-p_3-p)^3}\times(\frac{1}{p_1}+\frac{1}{p_2}-\frac{1}{p_3}-\frac{1}{p})\frac{-4}{\tau_c^3}\hspace{0.4cm}\big\}\ \Big]\notag\\
&+\frac{(-ig)^2}{6}\frac{1}{2p}\int\prod^3_{i=1}\frac{d^3p_i}{(2\pi)^32p_i}
\ (2\pi)^3\delta^{(3)}({\bf p}+{\bf p}_1+{\bf p}_2+{\bf p}_3)\notag\\
&\hspace{0.4cm}\times \Big[+3\{(1+f(p_1))(1+f(p_2))f(p_3)-f(p_1)f(p_2)(1+f(p_3))\}\notag\\
&\hspace{2.4cm}\times f(p_1+p_2-p_3)\ e^{+i(p_1+p_2-p_3)\bar{\tau}}\notag\\
&\hspace{2.4cm}\times\big\{
+\frac{1}{i(p_1+p_2-p_3-p)}\times\frac{1}{p^2}\frac{-2\bar{\tau}}{\tau_c^3}\notag\\
&\hspace{3.2cm}+\frac{1}{i(p_1+p_2-p_3-p)^2}\times(\frac{1}{p_1}+\frac{1}{p_2}-\frac{1}{p_3}-\frac{1}{p})\frac{-2\bar{\tau}}{\tau_c^3}\notag\\
&\hspace{3.2cm}+\frac{1}{(p_1+p_2-p_3-p)^2}\times(\frac{1}{p_1^2}+\frac{1}{p_2^2}+\frac{1}{p_3^2}+\frac{1}{p^2})\frac{-2}{\tau_c^3}\notag\\
&\hspace{3.2cm}+\frac{1}{(p_1+p_2-p_3-p)^3}\times(\frac{1}{p_1}+\frac{1}{p_2}-\frac{1}{p_3}-\frac{1}{p})\frac{-4}{\tau_c^3}\hspace{0.4cm}\big\}\ \Big]. \notag
\end{align}

Most of the IR divergences at $p_1+p_2-p_3-p=0$ cancel out between $C_{on}[f]$ and $C_{off}[f]$. 
This is because the total spectral weight is conserved due to unitarity. 
The remaining IR divergence comes from momentum dependence of the distribution function
\begin{equation}\begin{split}
f(p_1+p_2-p_3)=\ f(p)+f'(p)(p_1+p_2-p_3-p)+\cdots.
\end{split}\end{equation}
As explained in the main text, this IR divergence leads to the change of the distribution function
\begin{align}
\delta f\sim
&\ f'(p)\ \frac{g^2}{2}\frac{1}{2p}\int\prod^3_{i=1}\frac{d^3p_i}{(2\pi)^32p_i}
\ (2\pi)^3\delta^{(3)}({\bf p}+{\bf p}_1+{\bf p}_2+{\bf p}_3)\\
&\hspace{0.4cm}\times \Big[\{(1+f(p_1))(1+f(p_2))f(p_3)-f(p_1)f(p_2)(1+f(p_3))\}\notag\\
&\hspace{3.2cm}\times\frac{1}{(p_1+p_2-p_3-p)^2}(\frac{1}{p_1}+\frac{1}{p_2}-\frac{1}{p_3}-\frac{1}{p})\frac{2}{\tau_c^2}\ \Big]\notag. 
\end{align}
Here again we may adopt the IR cut-off : $|p_1+p_2-p_3-p|\sim 1/|\tau_c|$. 
$\tau_c$ dependence can be entirely absorbed into physical quantities $P_i=p_iH|\tau_c|$, $T=\beta H|\tau_c|$.

We may draw the following conclusion in this appendix. 
The leading order collision term is identical to that in Minkowski space. 
If we consider the higher order terms in $1/p|\tau_c|$ expansion, 
the off-shell part is generated due to the particle production while the total spectral weight is preserved 
due to unitarity. 
We further find the non-trivial change of the distribution function due to IR divergences. 
These features in $g\varphi^4$ theory are qualitatively identical to those in $\lambda\varphi^3$ theory. 

\section{Two and three point functions}
\setcounter{equation}{0}

\hspace{0.7cm}
The contribution to the three point function from the degrees of freedom
inside the cosmological horizon is evaluated as
\begin{align}
&\langle \varphi^3(x) \rangle\notag\\
=&\langle in|-i\frac{\lambda}{3!}\int^\tau_{-\infty}\frac{d\tau_1}{H^4\tau_1^4}\int d^3x_1 
[\varphi^3(x),\varphi^3(x_1)]\ |in\rangle\\
=& -\int_{|1/\tau|}\frac{d^3p}{(2\pi)^3}\frac{\lambda}{16\pi^2p^2}H^2\tau^2 \int_0^{\infty} d\varepsilon \notag\\
&\times\Big[\ \theta (\varepsilon - p)\int_{\varepsilon -p\over 2}^{\varepsilon +p\over 2} dp_1\notag\\
&\hspace{0.4cm}\left({1\over \varepsilon + p}\{(1+f(p))(1+f(p_1))(1+f(\varepsilon -p_1))-f(p)f(p_1)f(\varepsilon -p_1)\}
\right.\notag\\
&\hspace{0.4cm}\left.-{1\over \varepsilon -p}\{(1+f(p))f(p_1)f(\varepsilon -p_1)-f(p)(1+f(p_1))(1+f(\varepsilon -p_1))\}
\right)\notag\\
&\hspace{0.4cm}+2\theta (p-\varepsilon )\int_{p+\varepsilon \over 2}^{\infty}dp_1\notag\\
&\hspace{0.4cm}\left({1\over \varepsilon + p}\{(1+f(p))(1+f(p_1))f(p_1-\varepsilon)-f(p)f(p_1)(1+f(p_1-\varepsilon))\}
\right.\notag\\
&\hspace{0.4cm}\left.+{1\over p- \varepsilon}\{(1+f(p))f(p_1)(1+f(p_1-\varepsilon))-f(p)(1+f(p_1))f(p_1-\varepsilon)\}
\right)\Big].\notag
\end{align}
It in fact does not possess IR singularity around $\varepsilon-p\sim 0$.

We subsequently evaluate the contributions from the degrees of freedom outside the horizon.
We expand the original wave function in the power series of $p\tau$
\begin{equation}
\phi_{\bf p} (x) = -i{H\over \sqrt{2p^3}}(1+{1\over 2}(p\tau)^2-i{1\over 3}(p\tau)^3 \cdots )
e^{i{\bf p}\cdot{\bf x}} .
\end{equation}
We also assume $f=0$ in these estimates.

The expectation value of the three point function is 
\begin{align}
\langle \varphi^3(x) \rangle
=&\langle in|-i\frac{\lambda}{3!}\int^\tau_{-\infty}\frac{d\tau_1}{H^4\tau_1^4}\int d^3x_1\ 
[\varphi^3(x),\varphi^3(x_1)]\ |in\rangle\\
=&-i\lambda\int^\tau_{-\infty}\frac{d\tau_1}{H^4\tau_1^4}\int d^3x_1\prod^3_{i=1}\frac{d^3p_i}{(2\pi)^3}\notag\\
&\hspace{0.4cm}\times\big\{\phi_{{\bf p}_1}(x)\phi_{{\bf p}_2}(x)\phi_{{\bf p}_3}(x)\phi_{{\bf p}_1}^*(x_1)
\phi_{{\bf p}_2}^*(x_1)\phi_{{\bf p}_3}^*(x_1)-c.c.\big\}\notag\\
=&-\frac{\lambda H^2}{2^8\cdot 3}
\int\prod^3_{i=1}\frac{d^3p_i}{(2\pi)^3}(2\pi)^3\delta^{(3)}({\bf p}_1+{\bf p}_2+{\bf p}_3)\notag\\
&\times\Big[+\big\{
\left(\frac{p_1}{2}\right)^{-3}\left(\frac{p_2}{2}\right)^{-3}
+\left(\frac{p_2}{2}\right)^{-3}\left(\frac{p_3}{2}\right)^{-3}
+\left(\frac{p_3}{2}\right)^{-3}\left(\frac{p_1}{2}\right)^{-3}\big\}\notag\\
&\hspace{5.6cm}\times\int^\tau_{-1/max\{p_i\}} d\tau_1\ \frac{1}{-\tau_1}\notag\\
&\hspace{0.8cm}-\big\{
\left(\frac{p_1}{2}\right)^{-3}\left(\frac{p_2}{2}\right)^{-3}
+\left(\frac{p_2}{2}\right)^{-3}\left(\frac{p_3}{2}\right)^{-3}
+\left(\frac{p_3}{2}\right)^{-3}\left(\frac{p_1}{2}\right)^{-3}\big\}\notag\\
&\hspace{5.6cm}\times(-\tau)^{3}\int^\tau_{-1/max\{p_i\}} d\tau_1\ \frac{1}{(-\tau_1)^{4}}\ \Big]\notag\\
\approx&-\frac{\lambda H^2}{2^5\pi^4}
\int^1_{-\varepsilon_0\tau}\frac{dp_2}{p_2}\int^1_{-\varepsilon_0\tau}\frac{dp_3}{p_3}\int^1_{-1}d\cos\theta\ 
(-\log(max\{p_i\}))\notag\\
\approx&-\frac{\lambda H^2}{2^4\cdot 3\pi^4}(-\log(-\varepsilon_0\tau))^3.\notag
\end{align}
Here we have extracted only the leading term after rescaling $-p_i\tau\ \to\ p_i$. 

The leading perturbative correction to the two point function is 
\begin{align}
\langle \varphi^2(x)\rangle
=&\langle in|\left(\frac{-i\lambda}{3!}\right)^2
\int^\tau_{-\infty}\frac{d\tau_1}{H^4\tau_1^4}
\int^{\tau_1}_{-\infty}\frac{d\tau_2}{H^4\tau_2^4}\int d^3x_1d^3x_2\ 
\big[[\varphi^2(x),\varphi^3(x_1)],\varphi^3(x_2)\big]\ |in\rangle\notag\\
=&-\lambda^2
\int^\tau_{-\infty}\frac{d\tau_1}{H^4\tau_1^4}
\int^{\tau_1}_{-\infty}\frac{d\tau_2}{H^4\tau_2^4}\int d^3x_1d^3x_2
\prod^4_{i=1}\frac{d^3p_i}{(2\pi)^3}\\
&\times\big\{(\phi_{{\bf p}_1}(x)\phi_{{\bf p}_1}^*(x_1)-c.c.)
(\phi_{{\bf p}_2}(x_1)\phi_{{\bf p}_3}(x_1)\phi_{{\bf p}_2}^*(x_2)\phi_{{\bf p}_3}^*(x_2)
\phi_{{\bf p}_4}^*(x_2)\phi_{{\bf p}_4}(x)-c.c.)
\big\}\notag\\
=&-\frac{\lambda^2}{2^8\cdot 3^2}
\int\prod^4_{i=1}\frac{d^3p_i}{(2\pi)^3}
(2\pi)^6\delta^{(3)}({\bf p}_1+{\bf p}_2+{\bf p}_3)
\delta^{(3)}({\bf p}_4+{\bf p}_2+{\bf p}_3)\notag\\
&\times\Big[
-\big\{\left(\frac{p_4}{2}\right)^{-3}\left(\frac{p_2}{2}\right)^{-3}
+\left(\frac{p_4}{2}\right)^{-3}\left(\frac{p_3}{2}\right)^{-3}
+\left(\frac{p_2}{2}\right)^{-3}\left(\frac{p_3}{2}\right)^{-3}\big\}\notag\\
&\hspace{1.6cm}\times\int^\tau_{-1/max\{p_i\}}d\tau_1\ \frac{1}{-\tau_1}\int^{\tau_1}_{-1/max\{p_i\}}d\tau_2\ 
\frac{1}{-\tau_2}\notag\\
&\hspace{0.8cm}+\big\{\left(\frac{p_4}{2}\right)^{-3}\left(\frac{p_2}{2}\right)^{-3}
+\left(\frac{p_4}{2}\right)^{-3}\left(\frac{p_3}{2}\right)^{-3}
\big\}\notag\\
&\hspace{1.6cm}\times\int^\tau_{-1/max\{p_i\}}d\tau_1\ (-\tau_1)^{2}\int^{\tau_1}_{-1/max\{p_i\}}d\tau_2\ 
\frac{1}{(-\tau_2)^4}\notag\\
&\hspace{0.8cm}+\big\{
\left(\frac{p_2}{2}\right)^{-3}\left(\frac{p_3}{2}\right)^{-3}\big\}\notag\\
&\hspace{1.6cm}\times(-\tau)^{3}\int^\tau_{-1/max\{p_i\}}d\tau_1\ \frac{1}{(-\tau_1)}\int^{\tau_1}_{-1/max\{p_i\}}
d\tau_2\ \frac{1}{(-\tau_2)^4}\notag\\
&\hspace{0.8cm}+\big\{\left(\frac{p_4}{2}\right)^{-2\mu}\left(\frac{p_2}{2}\right)^{-2\mu}
+\left(\frac{p_4}{2}\right)^{-3}\left(\frac{p_3}{2}\right)^{-3}
+\left(\frac{p_2}{2}\right)^{-3}\left(\frac{p_3}{2}\right)^{-3}\big\}\notag\\
&\hspace{1.6cm}\times(-\tau)^{3}\int^\tau_{-1/max\{p_i\}}d\tau_1\ \frac{1}{(-\tau_1)^4}\int^{\tau_1}_{-1/max\{p_i\}}
d\tau_2\ \frac{1}{-\tau_2}\notag\\
&\hspace{0.8cm}-\big\{\left(\frac{p_4}{2}\right)^{-3}\left(\frac{p_2}{2}\right)^{-3}
+\left(\frac{p_4}{2}\right)^{-3}\left(\frac{p_3}{2}\right)^{-3}
\big\}\notag\\
&\hspace{1.6cm}\times(-\tau)^{3}\int^\tau_{-1/max\{p_i\}}d\tau_1\ \frac{1}{-\tau_1}\int^{\tau_1}_{-1/max\{p_i\}}
d\tau_2\ \frac{1}{(-\tau_2)^4}\notag\\
&\hspace{0.8cm}-\big\{\left(\frac{p_2}{2}\right)^{-3}\left(\frac{p_3}{2}\right)^{-3}\big\}\notag\\
&\hspace{1.6cm}\times(-\tau)^{6}\int^\tau_{-1/max\{p_i\}}d\tau_1\ \frac{1}{(-\tau_1)^4}\int^{\tau_1}_{-1/max\{p_i\}}
d\tau_2\ \frac{1}{(-\tau_2)^4}\ \Big]\notag\\
\approx&+\frac{\lambda^2}{2^5\cdot 3\pi^4}\int^1_{-\varepsilon_0\tau}dp_2\int^1_{-\varepsilon_0\tau}dp_3\int^1_{-1}
d\cos\theta\ 
\frac{1}{p_2p_3}\frac{1}{2}(\log(max\{p_i\}))^2\notag\\
\approx&+\frac{\lambda^2}{2^6\cdot 3^2\pi^4}(-\log(-\varepsilon_0\tau))^4,\notag
\end{align}
where $p_4=\sqrt{p_2^2+p_3^2+2p_2p_3\cos\theta}$.

\end{document}